\documentclass[preprint]{aastex}
\usepackage{graphicx}
\usepackage{natbib}
\begin{document}
\title{The three-dimensional structure of Saturn's E ring}
\author{M.M. Hedman$^{a,*}$, J.A. Burns$^{a,b}$, D.P. Hamilton$^{c}$, M.R. Showalter$^{d}$}
\affil{\it $^a$ Department of Astronomy, Cornell University, Ithaca NY 14853 USA \\
$^b$ Department of Mechanical Engineering, Cornell University, Ithaca NY 14853 USA \\
$^c$ Astronomy Department, University of Maryland, College Park MD 20742 USA\\
$^d$ SETI Institute, 189 Bernardo Dr., Mountain View CA 94043 USA\\
$^*$ Corresponding Author {\tt mmhedman@astro.cornell.edu} \\}

{\bf Abstract:} Saturn's diffuse E ring consists of
many tiny (micron and sub-micron) grains of water ice distributed between
the orbits of Mimas and Titan. Various gravitational
and non-gravitational forces perturb these particles' orbits, causing the ring's local particle density to vary noticeably with distance from the planet, height above the ring-plane, hour angle and time. Using remote-sensing data obtained
by the Cassini spacecraft in 2005 and 2006, we investigate the E-ring's three-dimensional structure 
during a time when the Sun illuminated the rings from the south at
high elevation angles ($>15^\circ$). 
These observations show that the ring's vertical thickness grows with
distance from Enceladus' orbit and its peak brightness density
shifts from south to north of Saturn's equator plane 
with increasing distance from the planet. These data also reveal a
localized depletion in particle density near Saturn's 
equatorial plane around Enceladus' semi-major axis. 
Finally, variations are detected in the radial brightness profile and the 
vertical thickness of the ring as a function of longitude relative 
to the Sun. Possible physical mechanisms and processes that may be responsible for some of these structures include solar radiation pressure, variations in the ambient plasma, and electromagnetic
perturbations associated with Saturn's shadow.

\pagebreak

\section{Introduction}

Saturn's E ring is a tenuous and diffuse
ring extending from the orbit of Mimas out to at least as far
as the orbit of Titan \citep{Srama06, Kempf06}.  Prior to Cassini's arrival at Saturn, 
analyses of Voyager and ground-based images 
had clearly demonstrated that the ring is 
composed primarily of extremely small ($<5 \mu$m ) grains and
that its vertically-integrated brightness 
peaks near Enceladus' orbit \citep{Show91}.  Earth-based
images also indicated that  the ring's vertical thickness 
increases with distance from the planet exterior
to Enceladus' orbit \citep{Show91, Nicholson96, dePater04}. These early data, reviewed
by \citet{BHS01}, already pointed towards Enceladus as the source of the 
E ring and the structure of this ring was used to explore 
the production, transport and loss of 
circumplanetary dust grains \citep{Horanyi92, Hamilton93, HB94, Juhasz02, Juhasz04, 
Juhasz07, Horanyi08}.

\nocite{Spencer09, Horanyi09}

Data from the Cassini Mission have spectacularly
confirmed that Enceladus is indeed the E-ring's source,
revealing a series of vents near the moon's
south pole that are launching  plumes of tiny ice grains
into orbit around Saturn (see Spencer {\em et al.} 2009 and references therein). However, the various  remote-sensing and in-situ measurements onboard Cassini
have also provided substantial new information about
the structure of the E ring itself (see Hor\'anyi {\em et al.} 2009 for a recent review). These data promise
to provide many new insights into the dynamics of this ring, 
as well as how the E-ring particles interact with Saturn's moons 
and magnetospheric environment \citep{Schenk11, Verbiscer07}. 

The currently published analyses of the Cassini E-ring
observations mostly utilize in situ data, and have 
revealed several interesting additional structures. 
Data from the plasma wave antennas show a depletion
in the local particle density near Saturn's equatorial
plane in the E-ring's core  \citep{Kurth06}. The dust detector onboard
the spacecraft has not clearly seen this feature, but
has confirmed that the vertical thickness of the ring
increases exterior to Enceladus' orbit \citep{Kempf08}. Furthermore,
data from this instrument indicate that the ring's vertical
thickness  also increases interior to Enceladus'
orbit, and that the peak particle density of the inner parts of the 
E ring occurs significantly south of Saturn's equatorial plane \citep{Kempf08}.

The extensive remote-sensing observations of the rings
obtained by Cassini not only confirm the existence of 
the above-mentioned ring features, but also
yield new insights into the ring's three-dimensional structure. 
In particular, edge-on images provide a more global view of the ring's 
vertical structure, and various observations reveal significant
azimuthal asymmetries in the ring.  Some of these features
can be understood in the context of existing models, while
others are unexpected and require further investigation.

Rather than attempt to provide a complete assessment 
of all the available remote-sensing data for the E ring,
this paper will focus on selected observations
that provide the best global views of the E-ring's large-scale
structure.  These observations were all made in 2005 and
2006, and thus provide a snapshot of the ring's 
structure at a particular  point in time when the Sun illuminated
the rings from the south at a fairly large elevation  
angle (19.8$^\circ$-15.8$^\circ$), and the planet's 
shadow only reached as far  as 213,000 km in Saturn's equator 
plane and thus did  not extend much beyond the inner edge 
of the E ring. Studies of the ring's time variability will therefore 
be the subject of a future work. Furthermore, by focusing 
on the data that best document the E-ring's
morphology, we also defer any detailed analysis of spectral
or photometric constraints on the ring's particle size distribution to 
a separate report (and see \citet{IE11} for a preliminary photometric analysis of
certain high-phase Cassini images). However, since non-gravitational, size-dependent  forces play an  important role  in shaping the E ring, we will
have to consider the typical particle sizes observed in the 
relevant images.
 
 We begin by discussing the imaging data used in this analysis 
and how it has been processed. 
We then describe a series of high signal-to-noise observations 
of the edge-on E ring and use those data to parametrize the ring's  vertical
structure. Next, we demonstrate that the
global structure of the ring does not vary much with longitude
relative to Enceladus, but does change significantly
with longitude relative to the Sun. Using multiple
data sets, we identify asymmetries in both the
ring's brightness and its vertical thickness. Finally,
we discuss some dynamical implications of these various 
observations, and the challenges they pose for future theoretical
modeling efforts.

\section{Observations}

While various remote-sensing instruments onboard Cassini have obtained
extensive data on the E ring, our investigation of the
E-ring's global structure will focus exclusively on data
obtained by the cameras of the Imaging
Science Subsystem
\citep{Porco04}, which provide the clearest pictures
of the ring's morphology. Furthermore, only data from
observations that are particularly 
informative regarding the global structure of the ring
will be used in this analysis. These include:
\begin{itemize}
\item  Multiple series of  nearly edge-on
images of the E ring obtained during Revs 17-23
(``Rev" designates a Cassini orbit around Saturn)
that were obtained in late 2005 through the middle 
of 2006. These observations are particularly useful
for determining the vertical structure of
the ring.  
\item A sequence of narrow-angle camera images taken along
the boundary of Saturn's shadow on the ring obtained during Rev 28
in late 2006. This sequence
provides information about azimuthal asymmetries
in the ring. 
\item The spectacular wide-angle camera mosaic of the Saturn system obtained
when Cassini flew through Saturn's shadow during Rev 28 in
September 2006. During this uniquely distant eclipse, Cassini
was able to image the rings at exceptionally high phase angles. This final
observation provides the highest signal-to-noise measurements
of the E ring and provides the most comprehensive view of the azimuthal
variations in the ring.
\end{itemize}
All these observations are from a relatively restricted time period (2005-2006). This is partially due to distribution of the available observations, but it also allows us to develop a model of the E-ring's structure at one particular point in time when the solar elevation angle was fairly large.
 
These remote-sensing observations only provide information 
about the particles that are sufficiently large to efficiently scatter 
light from the Sun into the camera. For light of a given 
wavelength $\lambda$, the scattering cross-section $\sigma$ of
a given particle depends on the particle's size (radius) $s$.  In the 
large-particle (geometrical optics) limit $\sigma\propto s^{2}$, while in the 
small-particle (Rayleigh scattering) limit, $\sigma \propto s^{6}$, and 
the transition between these two limiting trends occurs where 
$s \sim \lambda$ \citep{vandeHulst}. Spectral and photometric studies of the E ring  \citep{Show91, dePater04, IE11} and in situ measurements by the RPWS instrument \citep{Kurth06} and the High-Rate Detector (HRD) on the Cassini Dust Analyzer \citep{Kempf08} indicate that the E-ring's  particle size distribution is steeper than $s^{-4}$ for particles with $s > 1 \mu$m.  For such a steep size distribution, the particles with $s \simeq \lambda$ (i.e. 
the most common relatively efficient scatterers) will make the biggest 
contribution to the light scattered by the ring. The above-mentioned images
were made at wavelengths around 0.63 $\mu$m (0.42-0.97 $\mu$m for the 
high-phase mosaic), so the ring's appearance should be dominated by 
particles with radii between about half a micron and a micron.
Indeed, several of the observed structures in the ring can best be explained if the
typical observable ring particles are in this size range (see below).

The above images and the published in situ measurements probe overlapping
but different parts of the E-ring's particle size distribution. Specifically, the range of particle sizes observed by the remote-sensing data extends below the $s>0.9\mu$m limit for the published HRD data \citep{Kempf08}. The dynamics of E-ring particles  can vary dramatically with particle size, especially around 1 $\mu$m \citep{Horanyi92, HB94, Juhasz04, Juhasz07, Horanyi08}, so the sensitivity of the imaging data to sub-micron grains complicates comparisons between the remote-sensing and in situ observations. The contribution of sub-micron grains to the observed ring brightness is  uncertain because the particle size distribution on 
these scales is not yet well constrained in the E ring. 
For the Enceladus plume, measurements of the plasma environment by Cassini's 
Langmuir Probe suggest that the steep size distribution observed by RPWS and the 
HRD may extend through the sub-micron range \citep{Wahlund09, Yaroshenko09, 
Shafiq11}, and the CAPS instrument has even detected a significant number of
nanometer-scale grains within the plume \citep{Jones09}. However, the size 
distribution in the E ring will differ from that of the plume on account of the
larger particles re-impacting onto Enceladus \citep{Hedman09, Kempf10}, and
the smaller particles being rapidly eroded by sputtering from
charged particle impacts or lost by other dynamical mechanisms \citep{BHS01, Juhasz07}. Future analyses of in situ and
remote sensing data will likely provide better constraints on the sub-micron particle
size distributions, but for now we will simply urge the reader to not 
over-interpret the discrepancies between these remote-sensing observations and other data sets.

\section{Imaging data reduction}

All the relevant images were calibrated using the standard
CISSCAL routines \citep{Porco04} to remove instrumental backgrounds,
apply flatfields and convert the raw data numbers to $I/F$,
a standardized measure of reflectance that is
unity for a Lambertian surface at normal incidence and emission.
The images are geometrically navigated using the appropriate
SPICE kernels and this geometry was refined based on
the position of stars in the field of view. For the images
taken at sufficiently high ring-opening angles, this improved
geometry could be used to determine the E-ring's projected
brightness as a function of radius and longitude in
Saturn's equatorial plane. By contrast, the interpretation of
the nearly edge-on views of the rings required  additional
data reduction.

The most natural coordinate system to use with edge-on
images are the cylindrical coordinates $\rho, \theta$ and $z$,
where $z$ is the distance from Saturn's equatorial plane (positive
being defined as northwards), $\theta$ is an azimuthal coordinate, and
$\rho$ is the distance from the planet's spin axis. For any given
image, we can define a $\rho-z$ plane passing through Saturn's center
that is most perpendicular to the camera's line of sight\footnote{If the camera is at an azimuthal angle $\theta_C$, this plane corresponds to where $\theta=\theta_C\pm90^\circ$
in the limit where the range to Saturn approaches infinity.}, and then re-project the brightness
measurements onto a regular grid of $\rho$ and $z$ values. 
We have chosen to produce maps of the edge-on rings with resolutions
of 500 km in $\rho$ and 100 km in $z$ in order to
facilitate subsequent processing. For example, maps derived from different 
images can be easily combined to produce larger-scale mosaic
maps of the rings or to improve the signal-to-noise on the ring
(e.g., stars and moons in the images can be effectively
removed from the maps using a median filter prior to producing the
combined map).

The E-ring's optical depth is so low that all the particles within the camera's field of view are visible, hence the maps derived from edge-on images represent the integrated  brightness through the ring along the line of sight. Since the outer parts
of the ring appear superimposed on the inner parts, interpreting the observed brightness measurements is not straightforward. Fortunately, we can
remove these complicating projection effects using a 
deconvolution algorithm known as ``onion-peeling" 
\citep{Showalter85, Showalter87}. Such algorithms 
have been used previously to convert vertically-integrated brightness profiles of the 
edge-on E ring into estimates of the ring's normal $I/F$ 
as a function of radius \citep{dePater04}. However, our 
edge-on maps of the E ring have sufficient signal-to-noise 
that we can, for the first time, apply the algorithm directly to 
the two-dimensional maps themselves.

Applying an onion-peeling algorithm directly to $I/F$ data
yields a novel photometric quantity that has a similar relationship 
to the projected $I/F$ as the local particle number density has 
with optical depth. For a sufficiently low optical-depth ring 
consisting of spherical particles of size $s$, the observed 
optical depth  $\tau$ is given by the  following integral:
\begin{equation}
\tau=\int \pi s^2 N(l) dl,
\end{equation}
where $l$ is the distance along the line of sight and $N(l)$ is the
local particle number density in the ring. If these particles have an
albedo $w$ and a phase function $P(\alpha)$, then the ring's observed $I/F$ is
simply:
\begin{equation}
I/F=\frac{1}{4}wP(\alpha)\tau= \int \frac{1}{4}wP(\alpha)\pi s^2N(l) dl .
\end{equation} 
More realistically, the particles in the ring will have a range of sizes $s$, and a corresponding range of cross sections $\sigma(s)$, phase functions $P(\alpha,s)$ and albedos $w(s)$. 
Given the particle size distribution $n(s,l)$, we can integrate the above expression
over all particle sizes to obtain the following expression for the observed $I/F$:
\begin{equation}
I/F=\int \left[\int \frac{1}{4}w(s)P(\alpha,s)\sigma(s) n(s,l) ds\right] dl =\int (J/F) dl ,
\end{equation} 
where $J/F$ is something we call the  ``local brightness density''. This quantity 
has units of inverse length and can be regarded as the total effective surface 
area per unit volume in the ring. This parameter has the benefit of providing
a localized measure of the ring's light-scattering properties, which is much easier to interpret
than the integrated $I/F$.

The onion-peeling algorithm derives $J/F$ estimates from the $I/F$ data
using an iterative procedure. For each row in the maps (i.e., a fixed $z$), the algorithm 
starts with the pixel corresponding to the largest distance from Saturn (i.e., the largest $\rho$) and 
uses the measured $I/F$ to determine the $J/F$ in this outer fringe of the rings. 
This information is then used to estimate how much light this material should
contribute to the ring's measured brightness at smaller values of $\rho$.
Once this contribution to the observed $I/F$ has been removed from the remaining pixels
in the row, we can compute the local brightness density for the ring material
visible in the next closest pixel to the planet's spin axis. Iterating 
over all the pixels then yields $J/F$ as a function of $\rho$ at the $z$ 
of the selected row. 

Any instrumental background level in the $I/F$ maps will create spurious 
gradients in the peeled maps of $J/F$, so such backgrounds must be removed prior to applying 
the algorithm to a given map. In practice this is done by computing the
median brightness in the pixels more than 20,000 km from
Saturn's equatorial plane and subtracting this number from
the entire image. Also, since this inversion method 
involves repeated differencing of signals from different pixels, 
which tends to amplify any noise in the image (especially at low $\rho$), 
we re-bin the maps to a radial resolution of 5000 km/pixel prior 
to performing the inversion.

Once the maps have been suitably prepared, the actual onion-peeling
can be performed. For simplicity, the algebra used in this analysis assumes
the spacecraft is very far from the rings, which is a reasonable approximation
for all the images discussed here. These calculations also assume that 
the ring is azimuthally symmetric. This is not precisely true for the E ring 
(see below), but the brightness profiles derived under this assumption 
have shapes consistent with those derived from images with finite ring-opening
angles, so complicating the algorithm to account for azimuthal asymmetries
does not appear to be necessary at this point. 

Each iteration in the onion-peeling algorithm begins by estimating the brightness density
$J/F$ from the outermost pixel with nonzero signal in the $I/F$ map generated during the previous
iteration. Say this pixel covers the region between the  minimum and maximum  radii  
$\rho_1$ and $\rho_2=\rho_1+\Delta\rho$ and the vertical range 
$z_1$ to $z_2=z_1+\Delta z$, respectively. 
Since previous iterations of the algorithm have already removed the signal 
from all the material orbiting the planet at radii greater than $\rho_2$,
 the light seen in the pixel comes entirely from an annulus  extending between 
$\rho_1$ and $\rho_2$. Assuming the brightness density $J/F$ is constant within 
this annulus, then the observed $I/F$ is simply:
\begin{equation}
\frac{I}{F}=\frac{J}{F}\frac{V_o}{\Delta \rho \Delta z},
\label{jfeq}
\end{equation}
where $V_o$ is the volume of the part of the annulus that lies within the 
limits of the pixel:
\begin{equation}
V_o=\left[\rho_2^2\cos^{-1}(\rho_1/\rho_2) -\rho_1\sqrt{\rho_2^2-\rho_1^2}\right]\Delta z.
\end{equation}
Equation~\ref{jfeq} can thus be solved for $J/F$ in order to estimate
the average brightness in this range of $\rho$ and $z$. This number is
added to the appropriate pixel of the onion-peeled map of $J/F$.

This estimate of $J/F$ is also used to compute and
remove  the contribution of this ring annulus
to the other pixels in the $I/F$ map. Assuming the ring is azimuthally
symmetric, this annulus will contribute to pixels in the same row of the map 
(i.e., with the same values of $z_1$ and $z_2$). For any pixel covering the  
radial range between $\rho_i$ and $\rho_j=\rho_i+\Delta\rho<\rho_1$, the 
volume of the annulus captured in the pixel is:
\begin{equation}
V'=\mathcal{V}(\rho_i,\rho_2,\Delta z)-\mathcal{V}(\rho_j,\rho_2,\Delta z)
		-\mathcal{V}(\rho_i, \rho_1, \Delta z)+\mathcal{V}(\rho_j, \rho_1, \Delta z),
\end{equation}
where $\mathcal{V}(x,r,h)$ is the volume of the portion of a cylinder with radius
$r$ and height $h$ that lies exterior to a vertical plane that 
has a minimum distance $x$ from  the cylinder's symmetry axis:
\begin{equation}
\mathcal{V}(x,r,h)=\left[r^2\cos^{-1}(x/r) -x\sqrt{r^2-x^2}\right]h.
\end{equation}
Given $V'$, the contribution of the relevant annulus to the observed brightness
in the selected pixel is simply:
\begin{equation}
\frac{I'}{F}=\frac{J}{F}\frac{V'}{\Delta\rho\Delta z}.
\end{equation}
We can therefore subtract $I'/F$ from the signal in the appropriate
pixel of the $I/F$ map, and thereby remove the light scattered by this annulus from the
integrated brightness map. The map of the residual $I/F$
left over after this subtraction is performed on all relevant pixels
then forms the basis for the next iteration of the routine.

Of course, the outermost pixel in the map is always a special case,
since we cannot first remove the contribution due to ring material
exterior to this pixel.
However, in practice the brightness of the ring becomes sufficiently
faint near the edge of the frame that any residual signal has little effect
on the peeled map.

\section{Vertical structure of the E ring}

\begin{figure}
\resizebox{6in}{!}{\includegraphics{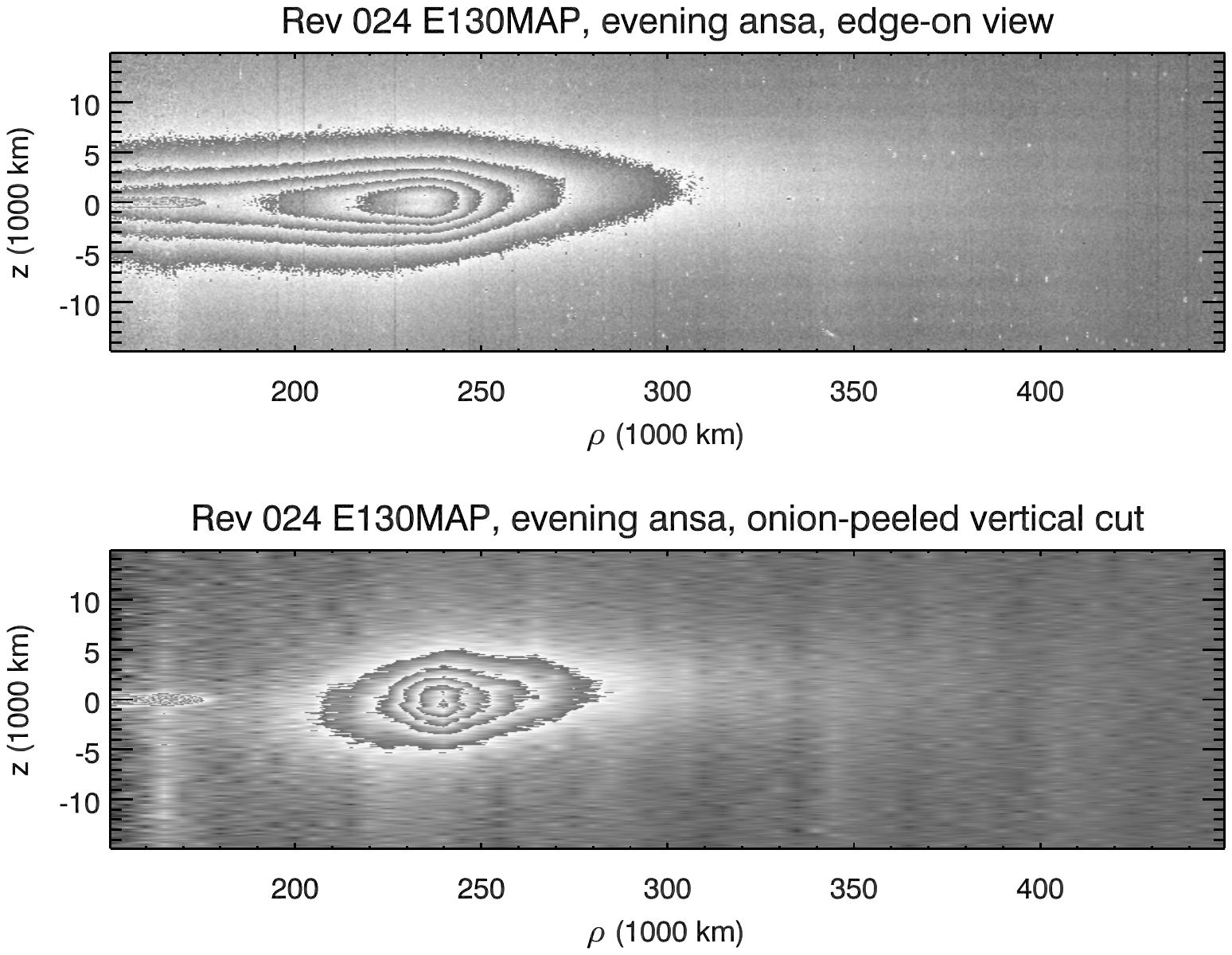}}
\caption{Edge-on views of the E ring derived from the Rev 024 E130MAP
observations. The top image shows the ring's  observed brightness as a function
of radius $\rho$ and $z$, while the bottom image shows the onion-peeled data, 
which represent the local brightness density as a function of $\rho$ and $z$.
For clarity, the vertical scale is enhanced by a factor of $\sim$3 relative to
the horizontal axis.
Both images use a ``cyclic'' stretch (where the displayed brightness
 equals the intrinsic ring brightness modulo some number) in order to 
 illustrate both the brighter and fainter parts
of the ring. A faint dip in brightness near the midplane can be seen near
the core of the ring around 240,000 km from Saturn center. Also note that the 
ring's peak brightness seems to occur at slightly positive values of $z$ 
outside 240,000 km and slightly negative values of $z$ inside 240,000 km 
in the onion-peeled image. 
}
\label{e130mapim}
\end{figure}

\begin{figure}
\resizebox{6in}{!}{\includegraphics{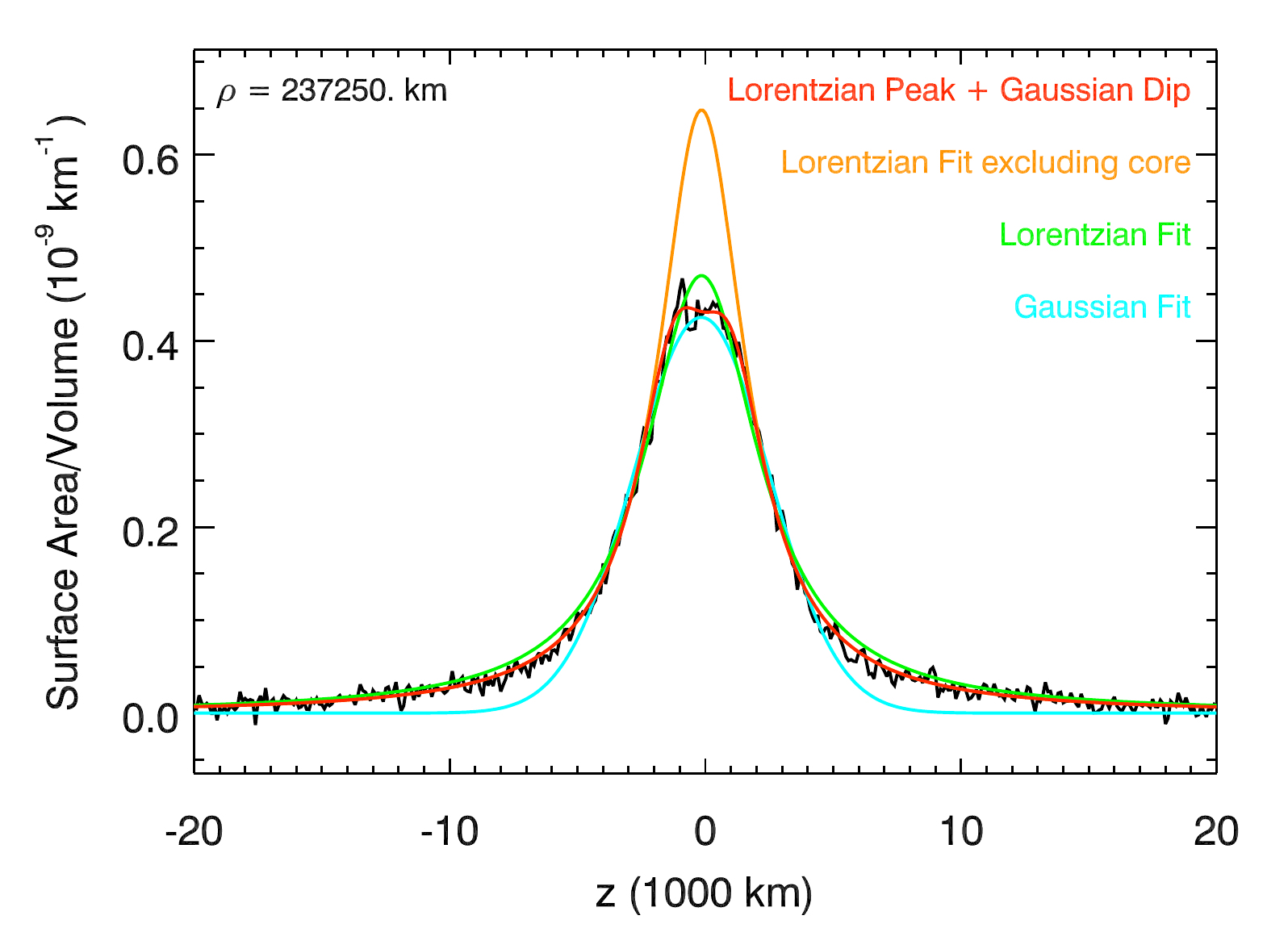}}
\caption{A sample vertical cut through the onion-peeled image derived
from the Rev 024 E130MAP data (Fig.~\ref{e130mapim}), together with various fits. The Gaussian
fit (in blue) clearly underestimates  the signal level at large values of $z$.
A simple Lorentzian fit (in green) does a much better job matching the overall
shape of the profile, but slightly overpredicts the flux at $z=0$  and in the
wings of the profile. This is largely due to the depletion in particle density 
near $z=0$. If we fit  only data with $|z|>$ 2000 km, we obtain the orange 
curve, which reproduces the wings of the profile better but grossly overpredicts the signal in the ring's core. Fitting the residuals between the data and the orange
curve to a Gaussian and combining the Lorentzian peak and the Gaussian dip
yields the red curve, which matches the data very well.}
\label{zprof}
\end{figure}

The best data set for documenting the detailed vertical structure of the E ring
is part of the E130MAP observation obtained on day 137 of 2006 in Rev 024. 
This is a large-scale
map of the E ring obtained at moderately high phase angles ($\sim130^\circ$)
and very low ring-opening angles ($\sim 0.1^\circ$), which yielded near edge-on,
high signal-to-noise data on the ring. A single map of the ring was constructed
from the longer-exposure WAC images on the lower-phase ansa of the ring,
which contained fewer stray light artifacts (W1526532467, W1526536067, 
W1526539667, W1526543267, W1526546867, W1526550467 and
W1526554067). While this map extends out to radii of 600,000 km, the
ring is only clearly detectable out to about 400,000 km.

Figure~\ref{e130mapim} shows the E-ring map derived from these data,
as well as a map of the local brightness density obtained by
onion-peeling the edge-on images. These maps clearly show the ring's
density peaks near Enceladus' orbit at 240,000 km, as expected.
What is perhaps more surprising is that the peak brightness does not
occur at $z=0$. Instead, there appear to be bands at $z=\pm1000$ km
that are slightly brighter than the region in between. While this ``double-banded''
structure is not apparent in data obtained by the HRD on the Cosmic Dust Analyzer
\citep{Kempf08, Kempf10}, it was previously detected in RPWS data \citep{Kurth06}.
We can also observe in  the peeled map that the peak of the ring's 
brightness occurs slightly north  of the ring-plane ($z> 0$) at radii outside 
the Enceladus' orbit, and slightly south of the ring-plane ($z<0$) interior to 
Enceladus' orbit. 

In order to better quantify the ring's vertical structure, we can fit vertical
brightness profiles derived from vertical cuts through the peeled map to various
functional forms, and plot the resulting fit parameters as functions of the radius $\rho$.
Figure~\ref{zprof} shows an example brightness profile and several different possible fits.
It is fairly clear that a Gaussian cannot accurately reproduce the observed vertical
profile. In particular, the E-ring profile has much broader wings than the best-fit
Gaussian model. By contrast, a Lorentzian fit does a much better job matching the
overall shape of the profile. However, even this fit is not perfect, as it overpredicts
the density in both the wings and around $z=0$. These discrepancies are almost certainly
due to the ``double-banded'' structure of the E-ring core discussed above. Even 
though the profile shown here does not have two distinct peaks separated by a 
local minimum\footnote{A true dip near $z=0$ is only clearly detected when a 
broader range of radii are co-added.}, it clearly has a flat-topped appearance that
reflects a localized depletion near the peak of the profile. If we exclude this region
and only fit the data at more than $2000$ km from the profile's peak, the resulting
model matches the shape of the wings very well, but grossly overpredicts
the signal near $z=0$. However, if we fit the residuals between this last model
and the data to a Gaussian, the resulting combination of a Lorentzian peak  and a Gaussian dip reproduces the observed profile remarkably well. It turns out that this 
combined model also works well in other parts of the E ring. 

\begin{figure}
\centerline{\resizebox{3.5in}{!}{\includegraphics{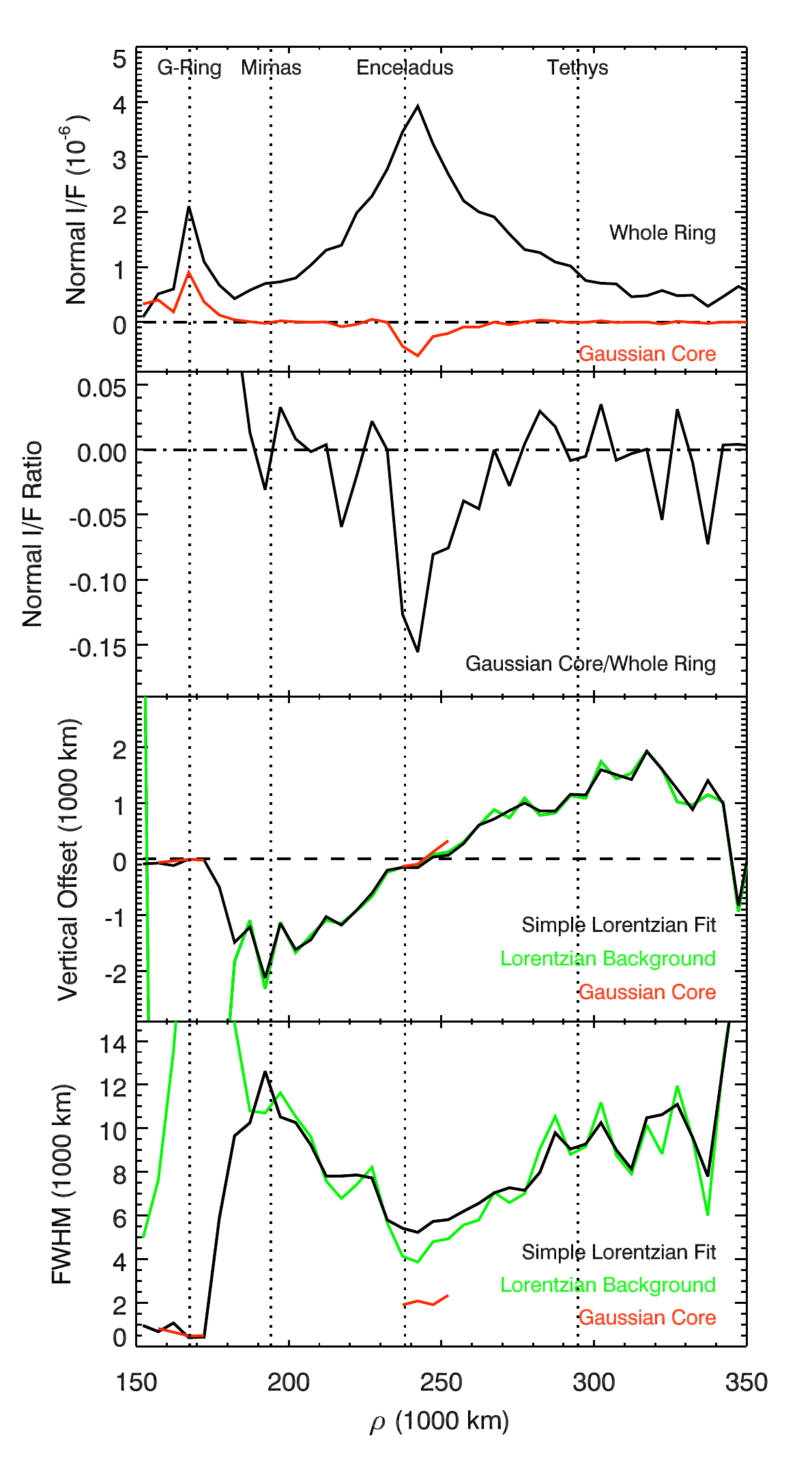}}}
\caption{Plots of the various fit parameters versus radius in the ring. The
top panel shows the normal $I/F$ of the entire ring derived by vertically
integrating the brightness measurements from the onion-peeled image. For comparison,
this panel also shows the integrated brightness of the residuals from the Lorentzian fit.
The second panel shows the ratios of the above two curves, which
is a measure of the magnitude of the depletion that produces the ``double-banded" structure. 
The vertical offsets of the peak location and the full-width at half-maximum 
of the various components of the fitted profiles are shown in the bottom two panels. In both 
cases, the black curves are the parameters for the single Lorentzian fit to the
entire profile, the green curves are the parameters for the Lorentzian fit
to the wings of the profile (more than 2000 km from the peak location), and the 
red curves are for the Gaussian fit to the residuals near the core of the ring. Note the 
signal to noise on these profiles falls dramatically outside 300,000 km, so the negative offsets near 345,000 km are not meaningful. Similarly, much of the
vertical structure interior to 200,000 km is strongly influenced by 
signals from the G ring.}
\label{parplot2}
\end{figure}

Figure~\ref{parplot2} shows the derived parameters as functions
of radius, including profiles of normal $I/F$ for the ring. These are simply
the integrals of the onion-peeled cuts over $z$, which correspond to the
$I/F$ that the ring would have if it were observed from exactly face-on. 
The ring's normal $I/F$  peaks near Enceladus' orbit, 
as expected based on previous profiles \citep{dePater04, Kempf08}. Also shown is the integral
of the residuals from the Lorentzian that best fits the wings of the vertical profile. This
can be regarded as a measure of the depletion near the equator plane 
responsible for the apparent double-banded structure, with more negative values
corresponding to a stronger signature. This feature is only detectable 
between 230,000 and 280,000 km, and has an asymmetric 
profile with a sharp inner edge and a more diffuse outer boundary. 
The depletion is also strongest near the orbit of Enceladus, and its
vertical full-width at half-maximum is 2000 km, which is comparable to
the width of the moon's Hill sphere \citep{Juhasz07, Kempf10, Agarwal11}.

Figure~\ref{parplot2} also shows the E-ring's vertical offset and width 
as functions of distance from the planet. The ring is centered at $z=0$ 
near the orbit of  Enceladus, and the vertical thickness of the 
ring is also at a minimum (4000-5000 km) here. 
The E-ring's peak brightness is displaced southwards of $z=0$ interior to 
Enceladus's orbit, getting as far as 1000-2000 km below 
Saturn's equator plane at the orbit of Mimas. Outside 
of Enceladus' orbit, the ring becomes displaced northwards, reaching
$z=+1000$ km near the orbit of Tethys (beyond which the signal-to-noise
in the images  degrades). The ring's vertical 
thickness  also increases roughly linearly with distance from Enceladus' 
orbit, rising to roughly 10,000 km at the orbits of Mimas and Tethys.  
All of these trends are roughly  consistent with earlier observations 
\citep{Nicholson96, dePater04, Kempf08}, except for the northward warp of
the outer ring, which had not been detected previously. The numerical
estimates of the ring's thickness, however, do diverge from those reported
in previous works.  Compared to the ground-based observations reported
in \citet{Nicholson96} and \citet{dePater04}, we obtain somewhat lower estimates
of the thickness near the core of the E ring. This is probably because
these earlier measurements were done on the projected $I/F$ data instead
of on peeled  maps, so their data from the E-ring core
include some contribution from the ring's flared outer regions. 

On the other hand, our thickness estimates are somewhat higher
than those derived from in-situ measurements. Using data from
the HRD on the dust detector, \citet{Kempf08} found that the ring's vertical 
full-width at half-maximum at the orbits of Mimas, Enceladus and Tethys are 
about 5400 km, 4300 km and 5500 km, respectively. A similar analysis of the RPWS data \citep{Kurth06}
indicated slightly thicker rings, with FWHM values ranging between 4000 and 5300 km 
near Enceladus' orbit, which is still somewhat lower than our estimates.
Some of the differences between the in-situ and remote-sensing estimates may be  attributed to azimuthal variations in the ring (see below) and differences in the assumed functional form of the E-ring's vertical profile, while others may reflect the different range of particle sizes probed by the various instruments (see Section 2). Smaller particles are expected to have a broader inclination distribution  \citep{Hamilton93, BHS01, Horanyi08}, and in-situ observations indicate that different-sized particles have different vertical scale heights in the ring \citep{Srama06}. The larger widths reported here may therefore be in part due  the remote-sensing data's finite sensitivity to smaller, sub-micron grains. However, detailed investigations of these issues is beyond the scope of this report.


\section{Longitudinal variations}

In the E ring, there are two physically meaningful  longitude 
systems, one measured relative to Enceladus and one measured
relative to the Sun. In principle, asymmetries could be
tied to either of these coordinate systems.

Let us first consider variations in the E ring that rotate around the
planet at the same rate as Enceladus. The most obvious of these
are the so-called ``tendrils", regions of enhanced optical depth 
near Enceladus that could either represent plume material that has 
not yet been dispersed into the E ring or local variations in the E-ring
density induced by the moon's gravitational perturbations on the
ring particle's orbits \citep{Spencer09}. These structures clearly follow Enceladus around
on its orbit. Furthermore, early ground-based observations detected
a bright cloud in the E ring that moved around the planet at roughly
the same orbital speed as Enceladus \citep{Roddier98}. While a 
detailed analysis of these localized structures is beyond the scope of 
this work, their existence
motivates us to explore the possibility that some of
the E-ring's  large-scale structure  may vary with longitude relative to
Enceladus.

A useful data set for examining such variations is the E80PHASE001 imaging sequence from Day 11 of 2006 in Rev 020. 
During this sequence, the cameras observed one ansa of the 
E ring for a period of 16 hours, or about one-half of Enceladus'
orbit period. 
During this time, the spacecraft stared at the same longitude
in the rings relative to the Sun as material rotated through.
We made maps of the 17 clear-filter images taken during the 
long stare that were targeted directly at the core of the E ring\footnote{Image names W1515651456, W1515652240, W1515656916,
W1515662987, W1515664339, W1515668736, W1515669520,
W1515674496, W1515680256, W1515686016, W1515691776,
W1515692560, W1515697536, W1515697810, W1515703308,
W1515704660 and W1515709057}. These
images were all obtained at moderate phase angles ($\sim 80^\circ$),
which meant that the ring was quite faint and spurious structures
in the images due to stray light scattered within the camera
\citep{West10} were more prominent. We were 
therefore unable to use the onion-peeling algorithm on these images,
and so only considered the edge-on observed brightness maps.
Fortunately, the stray-light patterns were stable from image to image, so
we could still use these data to look for variations with longitude
relative to Enceladus.

\begin{figure}
\centerline{\resizebox{3.5in}{!}{\includegraphics{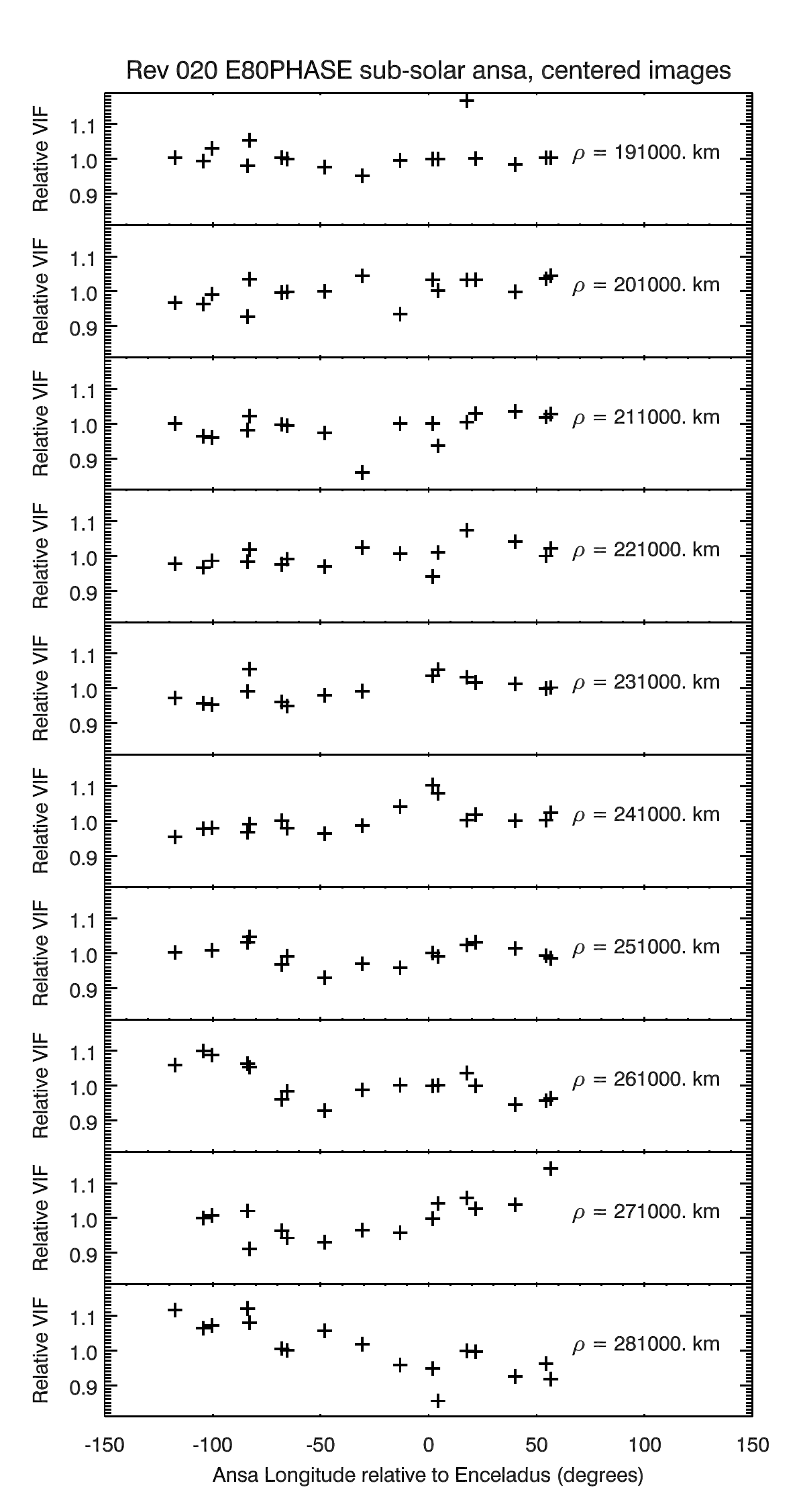}}}
\caption{Plots showing the vertically-integrated brightness of the edge-on E ring
as a function of longitude relative to Enceladus derived from the
Rev 020 E80PHASE data. All observations cover the same range of
longitudes relative to the Sun. Each panel shows the brightness
estimates normalized such that their average brightness is unity. 
Note that brightness measurements are not from onion-peeled maps,
so each panel includes contributions from material outside the labelled
distance. For 
all these plots, the brightness varies by less than 10\%, suggesting that
the brightness of the ring does not vary much with longitude relative
to Enceladus.}
\label{e8020lon}
\end{figure}

For each radius in the map derived from each image, 
we computed the vertically integrated brightness
of the ring within 15,000 km of Saturn's equatorial plane after
subtracting a cubic background based on the data at larger $|z|$. 
Figure~\ref{e8020lon} plots the variations in these brightness estimates with longitude 
relative to Enceladus. Note that the observed longitudes cover the region near Enceladus as well as both the leading and trailing Lagrange points, and thus these
data include most regions where one might suspect accumulations of material to be 
found. However, these plots only show subtle longitudinal brightness variations, many of which may simply be imaging artifacts. For example, there might be a slight (5-10\%) elevation in the ring's brightness in the vicinity of Enceladus' longitude at $\rho=241,000$ km, but this peak may simply be due to stray light from the moon itself. However, even if the interpretation of these variations are problematic, these plots clearly demonstrate that there are no systematic  brightness trends with
longitude at any radius within the ring above the 10\% level, which is much less than
the variations among observations made at different hour angles (see below).  Plots
of peak brightness and vertical width of the ring also fail to produce strong trends.
 We therefore conclude that,  despite the existence of localized tendril-like structures
in the vicinity of Enceladus, the broad-scale structure of the rings is not a strong function of longitude relative to Enceladus.

\section{Hour-angle variations}

The E-ring's large-scale structure varies much more dramatically with
longitude relative to the Sun than it does with longitude relative to Enceladus.
In order to avoid possible confusion regarding different longitude systems,
patterns that appear to track the Sun  will be referred to  as  ``hour-angle variations" or 
``hour-angle asymmetries''.  Multiple observing sequences document these asymmetries. First, selected edge-on views of the rings show asymmetries in the rings' brightness and vertical structure. Second, the data obtained during Rev 28, when Cassini flew through Saturn's shadow provides a global view of the E ring's asymmetric structure at very high  phase angles. Finally, images of Saturn's shadow within the E ring reveal further asymmetries within the ring's anti-solar quadrant.

\subsection{Edge-on views at moderate phase angles}

\begin{table}
\caption{Images used to construct maps from E105PHASE Imaging sequences}
\label{e105tab}
\resizebox{6in}{!}{\begin{tabular}{|c|c|c|c|c|c|} \hline
Rev 17 & Rev 18 & Rev 19 & Rev 22 & Rev 23 & Rev 23 \\
Sunward & Sunward & Shadow & Shadow & Sunward & Shadow \\ \hline
W1509652497 & W1512071423 & W1514471249 & W1521084669 & W1524468111 &	W1524509713 \\  
W1509652763 & W1512071689 & W1514473949 & W1521084967 & W1524474840 & W1524558146 \\
W1509659157 & W1512075983 & W1514475209 & W1521091512 & W1524481590 & W1524565192 \\
W1509667164 & W1512080543 & 			       & W1521098412 & W1524488340 & \\
W1509667978 & W1512085103 & 			       & W1521105312 & W1524495090 & \\
W1509674638 & W1512085369 & 			       & W1521112269 &			     & \\
W1509675718 & W1512089663 &			       &			  &			     & \\
W1509675984 & 			   & 			       &			  &			    & \\ \hline
 \end{tabular}}
\end{table}

\begin{figure}
\resizebox{6in}{!}{\includegraphics{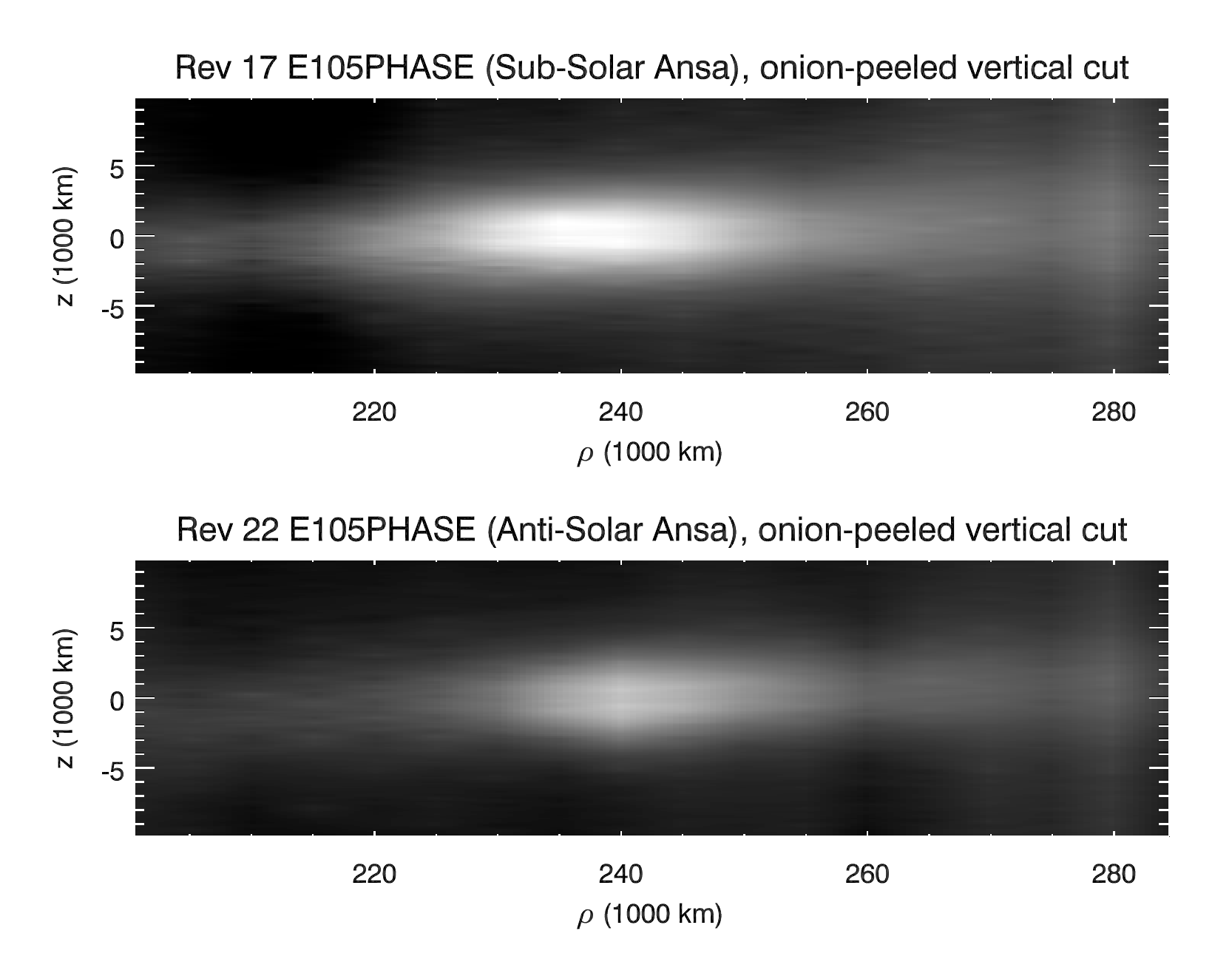}}
\caption{Sample onion-peeled maps of the E-ring's two ansa derived from E105PHASE
observations in Revs 17-23. Both images use the same stretch. Note the sub-solar
ansa is significantly brighter than the anti-solar ansa, and peaks at a smaller radius.
Also note that the double-banded structure is visible in both images.}
\label{e105phaseim}
\end{figure}

\begin{figure}
\centerline{\resizebox{4in}{!}{\includegraphics{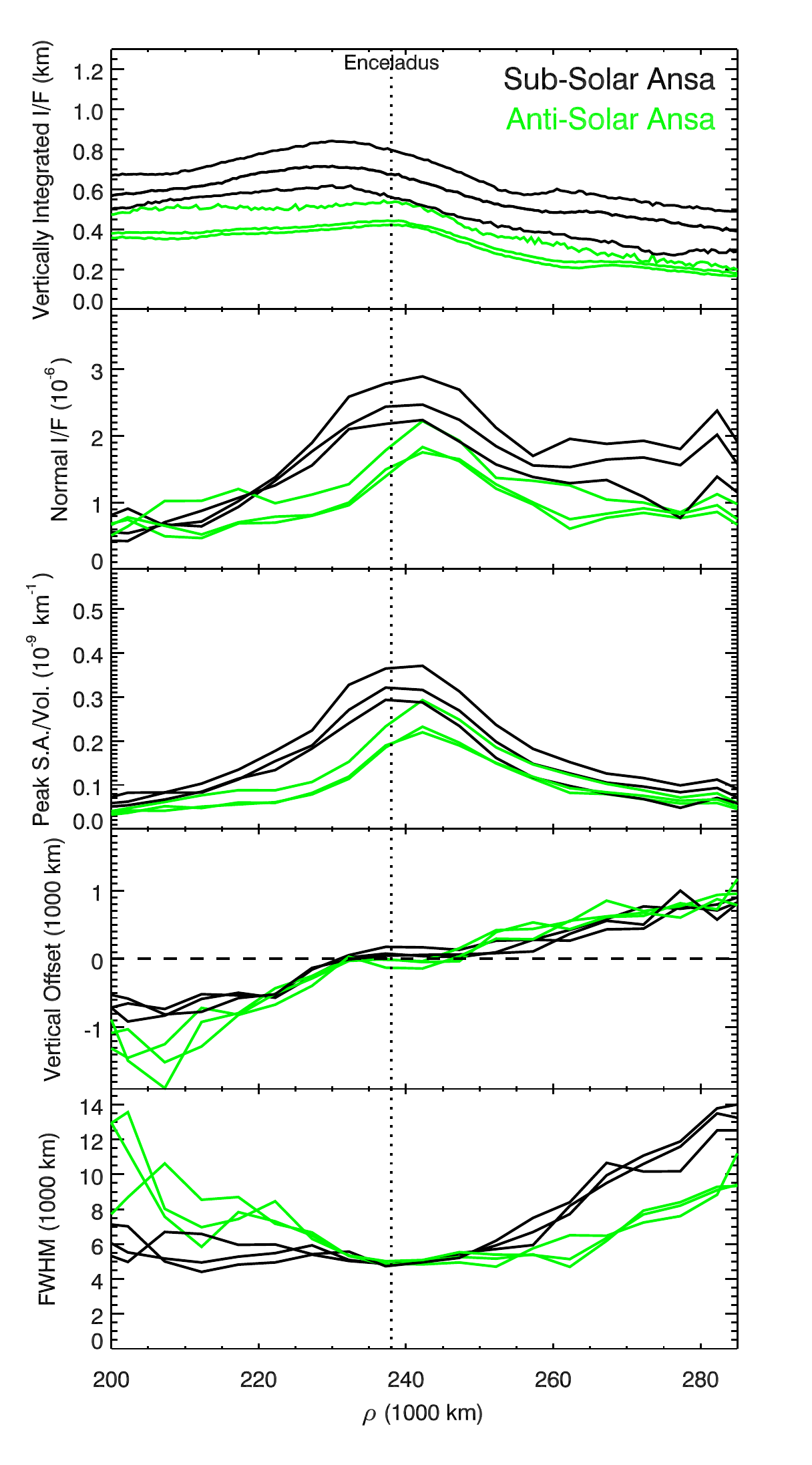}}}
\caption{Vertical structure of the sub-solar and anti-solar ansae of the 
E ring derived from clear-filter images obtained during E105PHASE observations. All results are derived from simple Lorentzian fits.
The top panel shows the vertically integrated brightness profiles of the ring. The second panel presents the normal $I/F$ derived by vertically
integrating the onion-peeled maps. The third panel displays the peak brightness ($J/F$) in the onion-peeled maps, while the bottom two panels show the vertical offset and
full width at half maximum of the ring. The black curves come from the 
sub-solar ansa while the green curves are for the anti-solar ansa. Note
that the inner part of the ring is systematically brighter on the 
sub-solar ansa (compare with Fig.~\ref{hpw177}). Also, while the vertical offset of
the ring appears to be similar on the two ansae, the vertical thicknesses are 
clearly different.}
\label{e105phasepar}
\end{figure}


During Revs 17-23 (days 306-2005 through 113-2006) Cassini observed the nearly edge-on E ring at  phase angles around $105^\circ$  in several separate E105PHASE campaigns. With this viewing geometry, one ring ansa is near the sub-solar longitude and the other is nearly aligned with the anti-solar longitude and Saturn's shadow. Note that when these observations were made, the solar elevation angle was above $17.7^\circ$, so Saturn's shadow does not extend much beyond 190,000 km from Saturn center within the E ring. Both ansae of the ring were viewed three separate times. The sub-solar ansa was observed during Revs 17, 18 and 23, while the anti-solar
ansa was observed during Revs 19, 22 and 23. A subset of the clear-filter images 
from each observation listed in Table~\ref{e105tab} was used to generate
sufficiently high signal-to-noise maps of the appropriate ansa for the onion-peeling
procedures described above. Examples of the onion-peeled maps of the sunward and shadowed ansa are shown in Figure~\ref{e105phaseim} with a common stretch applied to facilitate comparisons between the images. 
In both images the double-banded structure near the orbit of Enceladus is visible, but
we can also see differences between the images. In particular, the sub-solar ansa is clearly brighter than the anti-solar one, and the brightness peaks at a smaller radius. These trends are found in 
all the relevant edge-on maps and are consistent
with the high-phase observations described below.

In order to quantify these asymmetries, we fit vertical cuts through the peeled images
to simple Lorentzians (note the signal-to-noise ratios of these onion peeled maps were too low  for stable Gaussian+Lorentzian fits).  Figure~\ref{e105phasepar} plots the resulting fit parameters. Note that the parameters derived from the 
three observations of each ansa show the same trends, which means that the 
systematic differences between the two ansae are not due to instrumental artifacts. 
As before,  normal $I/F$ curves are derived by vertically integrating the brightness
measurements in the onion-peeled maps. In the sub-solar profiles, the normal $I/F$ peaks 
near Enceladus' orbit, while the anti-solar profiles peak about 1000 km
further out. Also, the sub-solar profiles are systematically brighter  than the
anti-solar profiles between 210,000 km and 250,000 km (and perhaps even further out  as well). These findings are consistent with those from the high-phase observations described below (see Fig.~\ref{hpw177}),  and demonstrate that this hour-angle asymmetry  is a stable and persistent 
feature of the E ring. 

Turning to the vertical structure of the ring, we find that the vertical offsets in the ring's
peak brightness density are essentially the same on the two ansae throughout the E ring, even between 210,000 and 250,000 km where the brightness asymmetries are most prominent. 
However, the trends in the ring's vertical thickness systematically differ for the two ansae (see Fig.~\ref{eringdraw} below for a cartoon sketch of these asymmetries).
Exterior to Enceladus' orbit, the E-ring's thickness grows more rapidly with increasing
radius on the sub-solar ansa than on the anti-solar ansa. On the other hand,
interior to Enceladus' orbit, the E-ring's thickness rises more steeply
away from Enceladus on the anti-solar side of the rings. These trends
are consistent among multiple observations and therefore cannot be ascribed
to instrumental artifacts. This asymmetry in the ring's vertical structure suggests
that those particles with their pericenters near the sub-solar longitude have a smaller range of inclinations than those whose pericenters are near the anti-solar point.

Unfortunately, Cassini did not obtain similarly  high signal-to-noise edge-on views of the ring
at extremely high or low phase angles, where the ansa would correspond to the  ring's ``morning" and ``evening''  sides (i.e., longitudes $\pm90^\circ$ away from the Sun).  
Both ansae were observed during the E130MAP observations described
above, but at these phase angles the ansae are halfway between the morning/evening
and sub-solar/anti-solar parts of the rings. These data show trends that are
qualitatively similar to those seen in the E105PHASE data, with the more sunward
ansa appearing brighter and having a more pronounced exterior flare than
the shadow-ward ansa. We therefore do not see any novel differences in the ring's vertical
structure  in these data that could easily be attributed to any asymmetries between the morning and evening sides of the ring.

\subsection{Extremely high-phase observations}

\begin{figure}
\resizebox{6in}{!}{\includegraphics{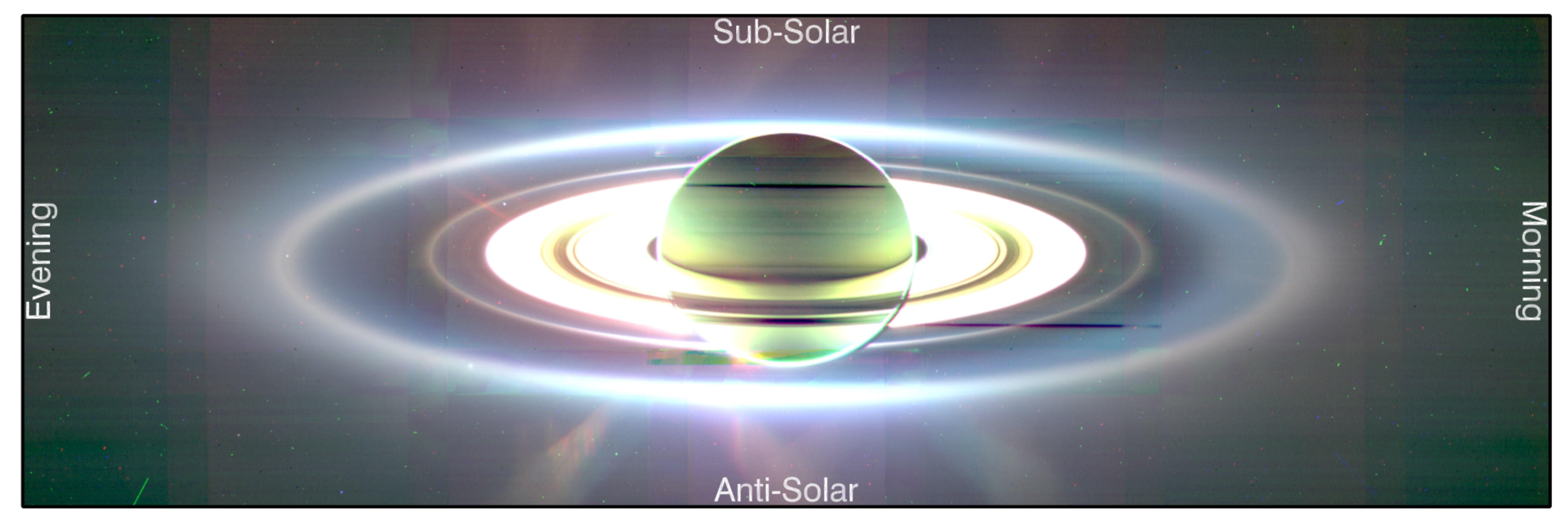}}
\caption{\small Mosaic of images obtained during the Rev 28 HIPHWAC sequence.
Red, green and blue in this image correspond to images taken in the
IR3, CLR and VIO filters, respectively, and a gamma correction has been applied to
make regions of differing brightness easier to see. The E ring is the outermost diffuse
ring in this image, while the G ring can be seen as a narrow feature between the diffuse E ring and the main ring system. Several camera artifacts can be seen in this image, including a horizontal line associated with image saturation, and diagonal bright streaks due to stray light from the planet.
Nevertheless, certain asymmetries in the E ring can also be observed in this mosaic.
The E ring is particularly bright where it passes near the planet because the ring is viewed  at higher phase angles that highlight small particles. Also the left, or evening, E-ring ansa has a similar color just inside and outside the E-ring core, while the right, or morning ansa appears bluer interior to the E-ring core than it does further from the planet. This blue region also appears to have a relatively sharp inner edge outside the G ring, which is not apparent on the opposite ansa. 
}
\label{hiphwac}
\end{figure}

\begin{figure}
\resizebox{6in}{!}{\includegraphics{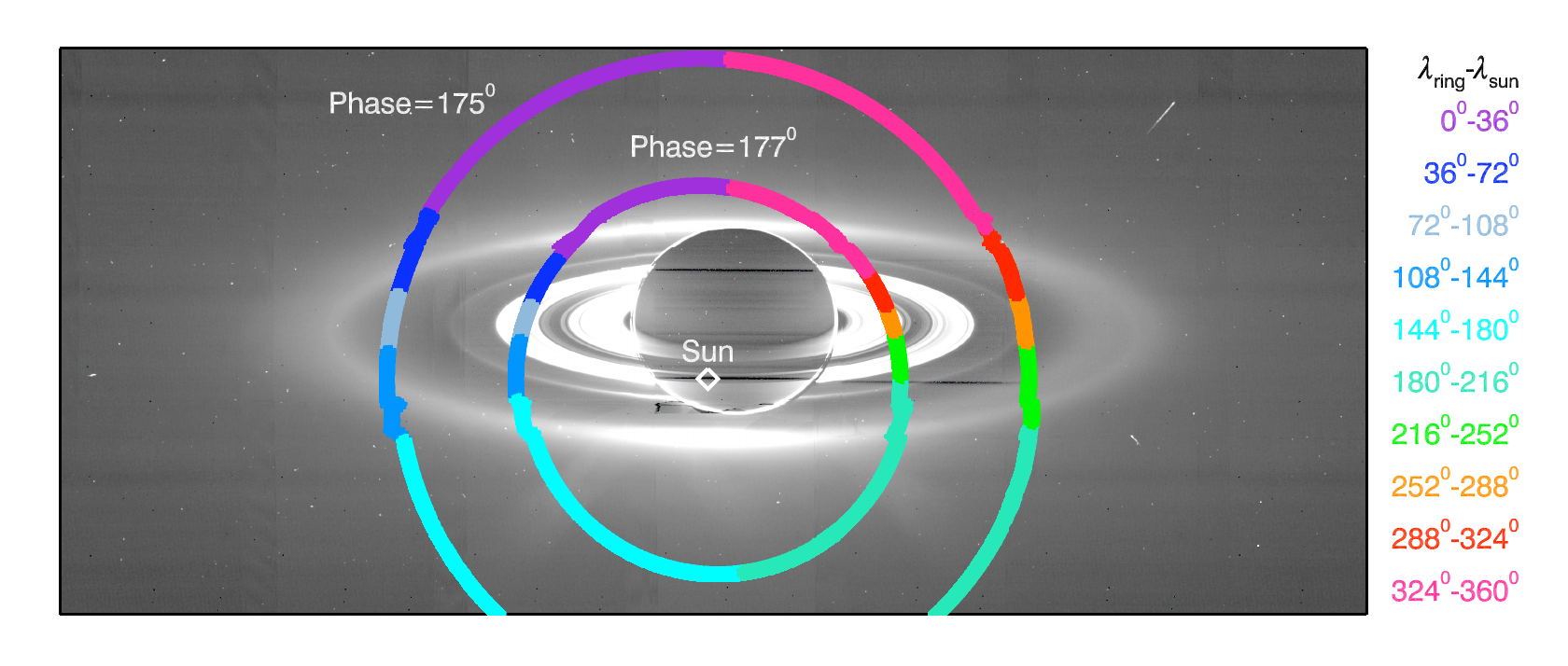}}
\caption{A graphical illustration of two lines of constant phase (175$^\circ$ and 177$^\circ$), overlaid on the VIO filter image mosaic shown in Figure~\ref{hiphwac} (also gamma-stretched). The colors along each circle indicate the longitude of the observations relative to the Sun, using the same color codes as Figures~\ref{hpw177} and~\ref{hpw175}. Slight ripples in the two circles occur at the boundaries between images within the mosaic due to small variations in the viewing geometry over the course of the observation.}
\label{hiphov}
\end{figure}

Figure~\ref{hiphwac} shows
a false-color mosaic derived from images (W1537023863-W153703340)
taken through the IR3, CLR and VIO filters (effective central wavelengths
420 nm, 634 rm and 917 nm, respectively)  while Cassini flew through Saturn's shadow on Day 258 of 2006 during the Rev 28 HIPHWAC observation.
Note the Sun is located behind the
planet in this image (a bright spot at about 7 o'clock on Saturn's limb hints at 
its location). The rings in the 
upper half of the mosaic are closer to the Sun,
while those in the lower half are closer to Saturn's shadow.
The left ansa in this image is thus the evening side of the rings
and the right ansa lies on the morning side of the rings. 
During this observation, Enceladus was moving from the evening ansa
towards the anti-solar side of the rings.
This mosaic contains several instrumental artifacts, including a
dark horizontal band due to partial saturation of the detector, 
and various bright streaks and flares due to stray light from the planet.
Nevertheless, one can still identify
several asymmetries in the ring's brightness and color from a visual
inspection of this image.

Perhaps the most obvious E-ring asymmetry is the difference
in the color of the two ansae.  On the evening ansa, the ring has a similar
color just interior and exterior to the E-ring core, while on the morning
ansa, the inner parts of the ring appear distinctly ``bluer'' than the outer parts.
Also, on the evening ansa the ring's brightness smoothly decreases inwards
towards the G ring, while the morning ansa seems to have a sharper inner
edge somewhat exterior to the G ring.
In addition, the sub-solar and anti-solar sides of the rings show a
more subtle, but still important, asymmetry.
In this mosaic, the Sun was located somewhat below 
the center of the planet, 
so the anti-solar side of the rings is being observed at a somewhat 
higher phase angle than the sub-solar side of the rings. 
Since a dusty ring's brightness should increase with phase angle \citep{Show91, BHS01}, 
the anti-solar part of the rings should be brighter than the sunward region, but in fact
these two sides of the ring actually appear to be equally bright,
which implies that the E ring is intrinsically brighter
near the sub-solar longitude\footnote{\citet{IE11} came to the same conclusion
in an independent analysis of these images}. The edge-on images discussed
above are consistent with this interpretation, and furthermore the
fact that those observations always show the sub-solar ansa to be brighter
than the anti-solar ansa means that this asymmetry is a persistent ring feature and
not a result of a coincidence involving the location of Enceladus in these images.

In order to explore these asymmetries more quantitatively, 
we must generate profiles of brightness versus radius. However,
to be informative, these profiles need to account for the significant changes
in the phase and emission angles across this mosaic. The emission-angle 
variations can be dealt with relatively simply by computing
the emission angle at the ring-plane for each pixel in the images and
multiplying the observed $I/F$ by the cosine of the emission angle
$\mu$ to obtain the normal $I/F$, i.e., the predicted brightness
of the ring when it is viewed at normal incidence. For low-optical-depth
rings like the E ring, this quantity should be independent of the
observed emission angle. 

\begin{figure}
\centerline{\resizebox{3.5in}{!}{\includegraphics{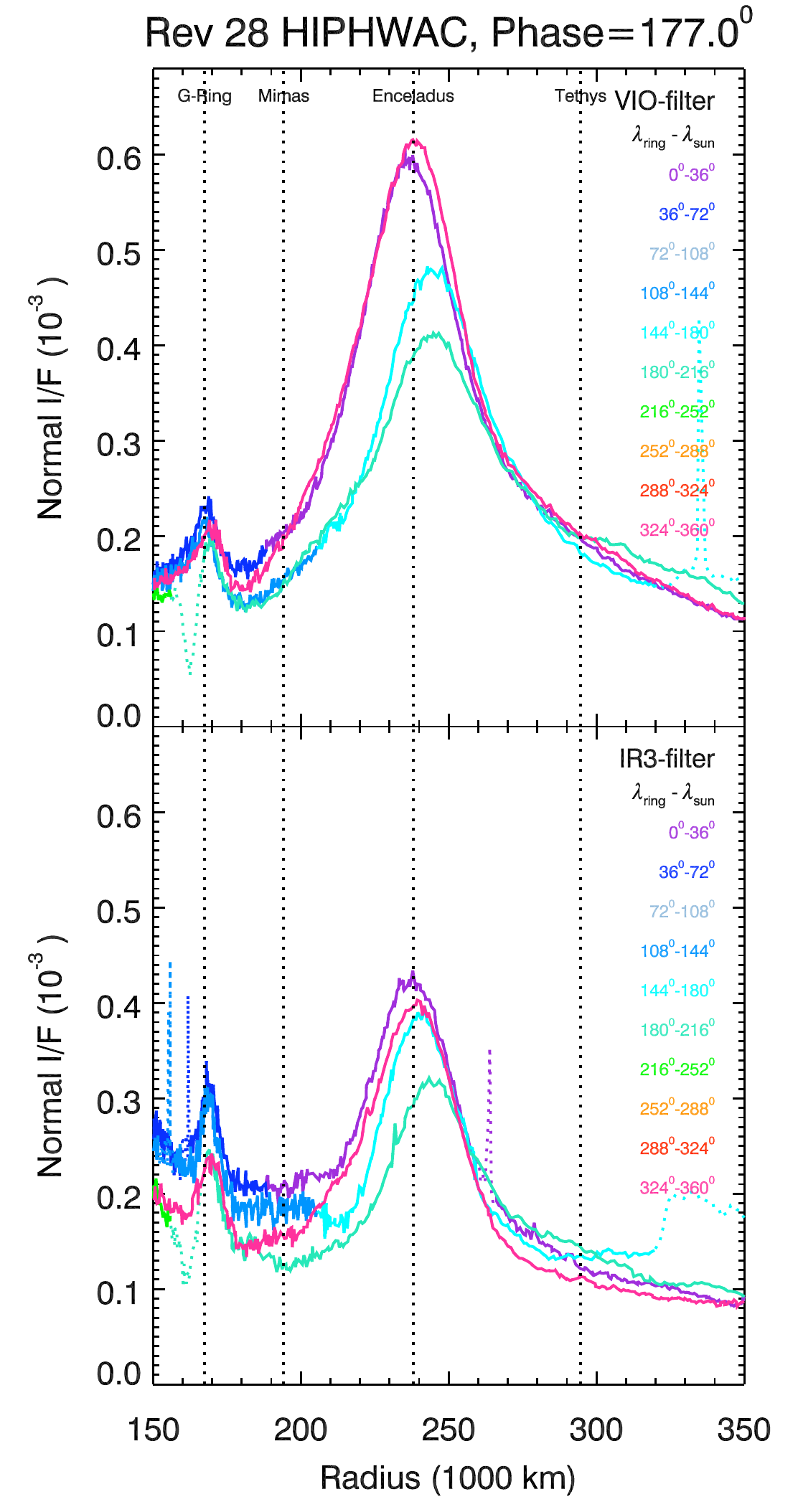}}}
\caption{Radial brightness profile derived from images of the E ring obtained
during the Rev 28 eclipse observations. All data come from a narrow range
of phase angles within $\pm0.05^\circ$ of 177$^\circ$, and are color-coded
by longitude relative to the Sun (see Fig.~\ref{hiphov}). The top panel shows data from
VIO-filter (420 nm) images and the bottom panel shows data from IR3-filter (917 nm) images. Dotted lines in the profiles indicate regions of the profile corrupted by instrumental artifacts: narrow positive spikes are generated by stars or cosmic rays, while  the narrow dips around 160,000 km are likely due to a change in the instrument's response along lines that became saturated on  the bright parts of Saturn's limb, and the plateau around 320,000 km
is due to stray light from the planet.  
Note that in the VIO-filter images the region
between 190,000 and 250,000 km is systematically brighter on the sub-solar
side ($\lambda-\lambda_\sun \simeq 0^\circ$) of the rings.}
\label{hpw177}
\end{figure}

\begin{figure}
\centerline{\resizebox{3.5in}{!}{\includegraphics{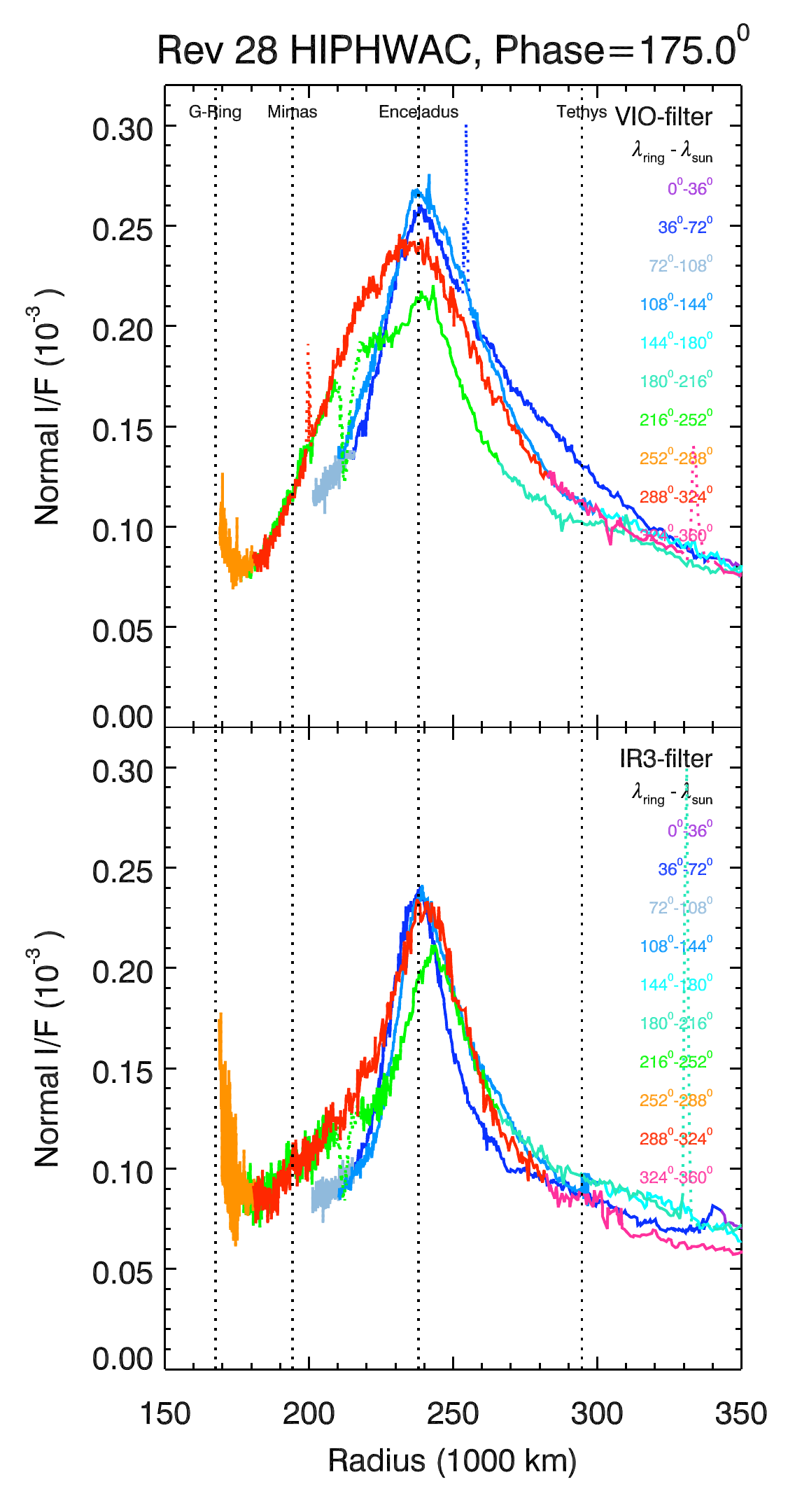}}}
\caption{Radial brightness profiles derived from E-ring images obtained during
the Rev 28 eclipse observations. All data come from a narrow range
of phase angles within $\pm0.05^\circ$ of 175$^\circ$. The data points are color-coded
by longitude relative to the Sun (see Fig.~\ref{hiphov}). The top panel shows data from VIO-filter (420 nm) images and the bottom panel shows data from IR3-filter (917 nm) images. Dotted lines in the profiles indicate regions  corrupted by instrumental artifacts: positive spikes are generated by stars or cosmic rays, while  the narrow dips around 210,000 km are likely due to a change in the instrument's response along lines that became saturated on bright parts of Saturn's limb. 
Note the differences in the shape of the VIO-filter profiles on the morning ($\lambda-\lambda_\Sun\sim270^\circ$) and evening ($\lambda-\lambda_\sun\sim90^\circ)$ ansa.}
\label{hpw175}
\end{figure}

Dealing with the phase-angle variations is more difficult, because 
we do not know {\it a priori} how the ring's brightness changes with
phase angle, especially if there are azimuthal asymmetries in the ring. 
Fortunately, we can avoid this issue by constructing brightness profiles at a fixed
phase angle. Lines of constant
phase angle correspond to circles in the mosaic centered on the apparent
position of the Sun behind the planet, with the circle's radius being set 
by the phase angle (see Fig.~\ref{hiphov}). For phase angles between 174.5$^\circ$
and 177.5$^\circ$, these circles cut through the E-ring's core at four separate
longitudes. Figs.~\ref{hpw177} and~\ref{hpw175} show the profiles corresponding 
to phase angles of 177$^\circ$ and 175$^\circ$, respectively (these phase angles were chosen
to minimize stray-light contamination within the core of the E ring). In both figures,
two sets of profiles are shown, one derived from VIO-filter
images (effective wavelength 420 nm) and the other derived from IR3-filter images (effective  wavelength 917 nm). In both cases, only data from the images
with relatively short exposures were used to avoid saturation effects.

First,  consider  the brightness
measurements made at $177^\circ$ phase in Fig.~\ref{hpw177}. The cuts through the E-ring's
core now fall either close to the sub-solar
side of the rings ($\lambda-\lambda_\Sun \sim 0 ^\circ$, purple-pink in Fig.~\ref{hpw177}) or the anti-solar side of the rings 
($\lambda-\lambda_\Sun\sim 180^\circ$, green-blue in Fig.~\ref{hpw177}). The IR3-filter data do not show strong azimuthal variations; the offsets between the profiles
primarily reflect the different instrumental background levels
seen interior to the ring. However, the VIO-filter profiles show clear asymmetries. In this case, 
the peak brightness occurs near Enceladus'  orbit on the sub-solar side
of the rings, while on the anti-solar side the peak lies about 10,000 km further from 
Saturn. Furthermore, the peak brightness on the sub-solar side 
is roughly 50\% higher than the peak brightness on the anti-solar side.
Thus, while outside 250,000 km the profiles are basically the same,
between 190,000 and 250,000 km the sub-solar ring is always brighter than
the anti-solar ring in the VIO-filter images. This asymmetry in the rings' brightness
is consistent with that seen in the edge-on views described above (compare Fig.~\ref{e105phasepar}).

Next, consider the 175$^\circ$ phase data in Fig.~\ref{hpw175}, where the cuts through the ring's core
fall near the morning ansa ($\lambda-\lambda_\Sun \sim 270^\circ$,  orangish-red and green in Fig.~\ref{hpw175}) or the  evening ansa ($\lambda-\lambda_\Sun \sim 90^\circ$, various shades of blue in Fig.~\ref{hpw175}). 
The brightness profiles derived from the IR3-filter images
are basically the same at the two ansa, but the VIO-filter  shows a clear asymmetry. 
The evening-ansa, VIO-filter profiles have a slightly wider peak
than the IR3 profiles (which is why the core of the ring in that part of the mosaic
has a reddish core and blue wings in Fig.~\ref{hiphwac}), but otherwise the IR3-filter and
evening-ansa VIO-filter profiles have similar shapes with a rather sharp 
peak at Enceladus' orbit. By contrast, the morning-ansa VIO-filter
profiles maintain a higher brightness (relative to the peak at Enceladus'  orbit) between 
210,000 and 240,000 km, which corresponds to the bluish-colored
region on the morning ansa that can be seen in Fig.~\ref{hiphwac}. Around 210,000 km, there is a relatively sharp drop in brightness that corresponds to the edge of this region in the mosaic. Further information about this morning-evening asymmetry can be obtained by examining high-phase images of the region around the edge of Saturn's shadow.

\subsection{High-phase shadow-edge observations}

\begin{figure}[tbp]
\resizebox{3in}{!}{\includegraphics{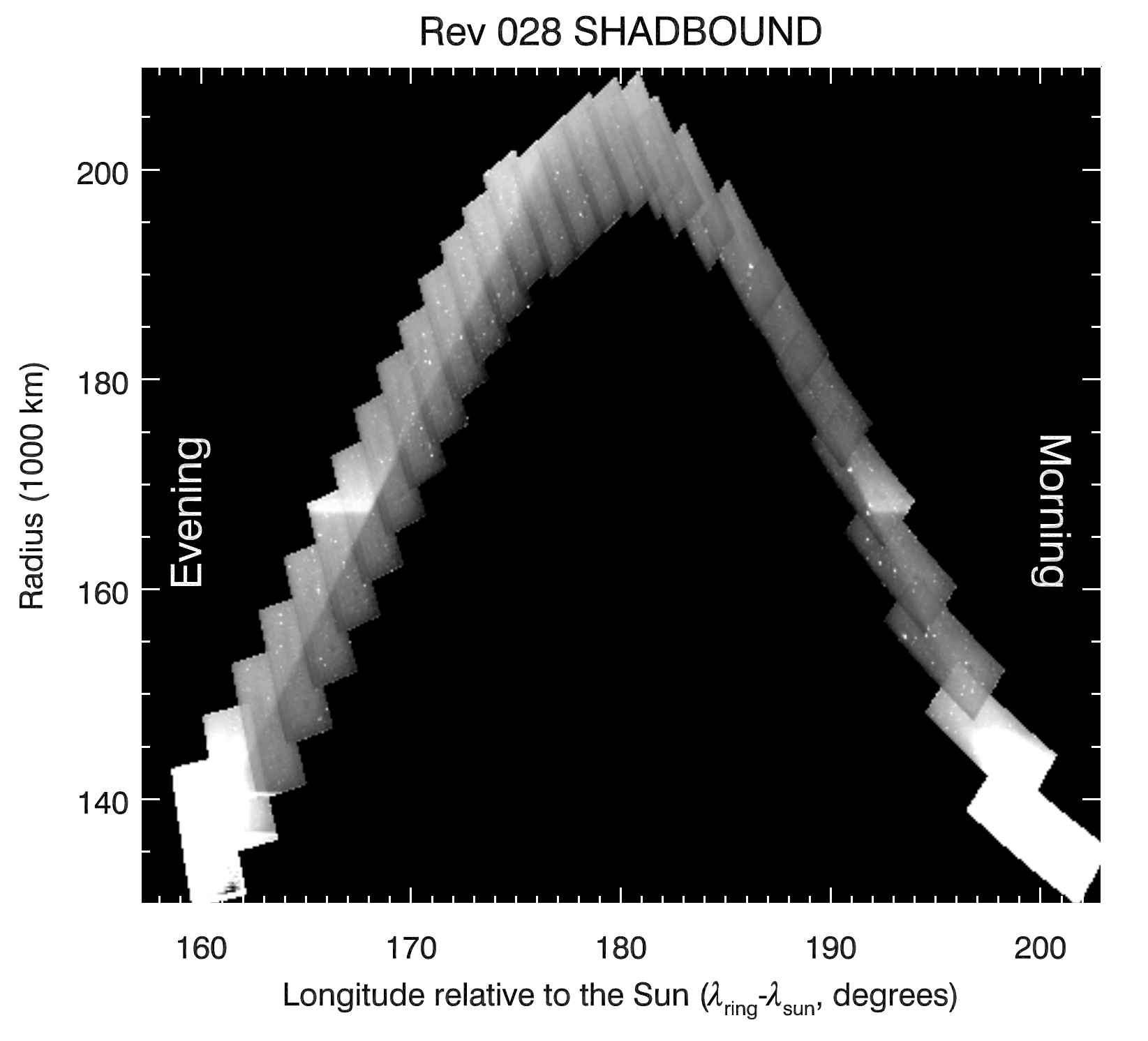}}
\resizebox{3in}{!}{\includegraphics{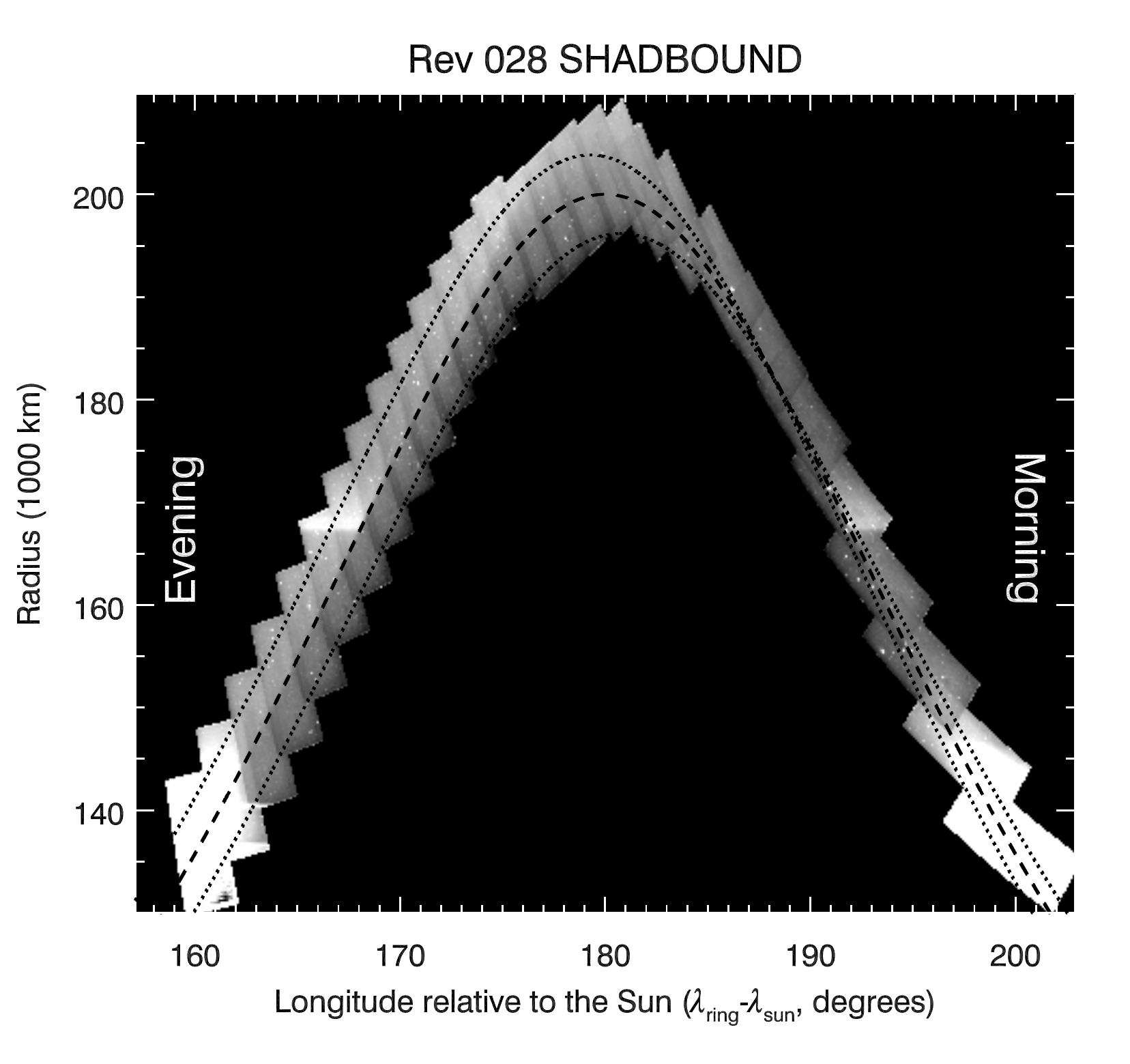}}
\caption{A mosaic of images obtained during the Rev 028 SHADBOUND sequence.
These images trace out the arc of Saturn's shadow on Saturn's equatorial plane (indicated by the dark dashed line in the right-hand plot) and have been re-projected onto a regular grid of radii and longitudes relative to the Sun. The saturated ends of the mosaic correspond to the edge of the A ring, and the G-ring can be seen outside the shadow around 167,500 km. Note that the brightness contrast across the shadow edge is much more prominent along the evening edge of the shadow than it is on the morning side (the G ring also appears somewhat brighter along the evening side, but this is largely an artifact of the background levels, see Fig.~\ref{shadboundprof}). In the right-hand mosaic, the dashed line indicates the position of Saturn's shadow in the planet's equatorial plane, while the two dotted lines mark the apparent positions of Saturn's shadow in planes at $z=+2000$ km (lower curve) and $z=-2000$ km 
(upper curve) }
\label{shadboundim}
\end{figure}

\begin{figure}[tbp]
\resizebox{6in}{!}{\includegraphics{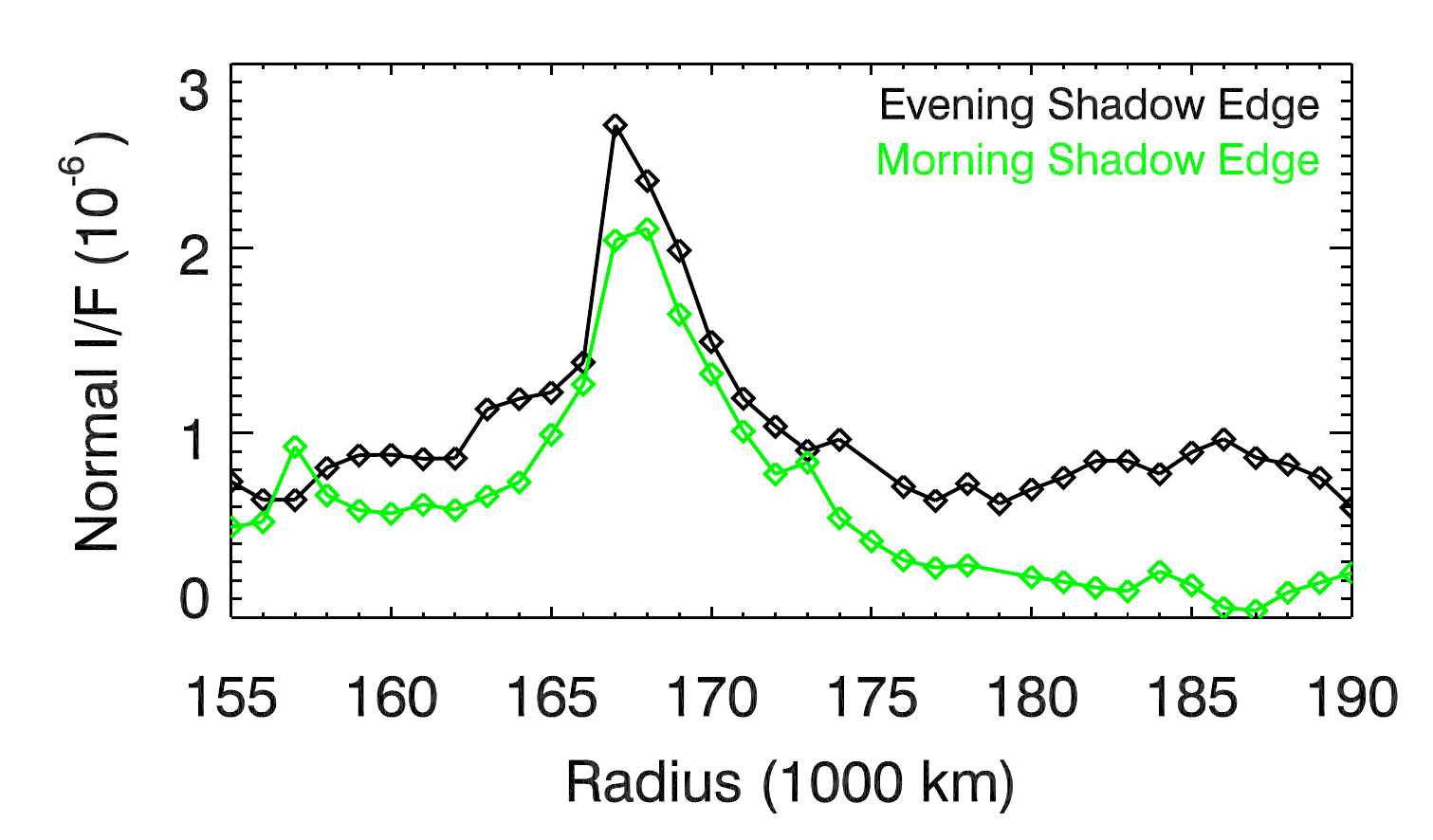}}
\caption{Profiles of the ring brightness versus ring-plane radius along the two edges of the shadow. Each profile is a plot of the brightness difference between regions 0.2$^\circ$-0.4$^\circ$ on either side of the predicted shadow position, which eliminates instrumental backgrounds. Each plotted brightness value is an average of  multiple measurements derived from individual rows or columns of the relevant images (whether rows or columns are used depends on which is more perpendicular to the shadow edge). These individual brightness measurements are median filtered prior to averaging to remove contamination from stars and cosmic rays. The peak at 167,000 km in both profiles is due to the G ring. The brightness level exterior to the G ring is systematically higher on the evening side of the shadow, consistent with Fig.~\ref{shadboundim}. }
\label{shadboundprof}
\end{figure}

Brightness asymmetries in the E ring can also be identified
in  a mosaic of images (N1536644151-N1536651126)
obtained during the Rev 28 SHADBOUND observation on Day 254 of 2006 at phase angles between  145$^\circ$ and 150$^\circ$ (see Fig.~\ref{shadboundim}). These images trace out the edge of Saturn's shadow on the planet's equatorial plane, and
indeed the shadow edge is visible in many of the images, especially near the G ring. 
Differencing the observed signals interior and exterior to the shadow edge enables us to remove any sky or instrumental backgrounds and thus provides estimates of the ring's absolute brightness near the two sides of the shadow (see Figure~\ref{shadboundprof}). 

There is a clear asymmetry in the brightness contrast along the two sides of the shadow, which is most prominent exterior to the G ring at 167,500 km.
The difference in the normal $I/F$  across the evening edge of the shadow is 
of order $10^{-6}$ between 170,000 and 190,000 km, while on the morning edge of the shadow the normal $I/F$ change is 5-10 times smaller.  Thus the E-ring is about an order of magnitude fainter on the shadow's  morning side 
 than it is on the evening side, far more than would be expected
 given slight differences in the phase angle between these two regions
 \citep{Show91}.

This difference between the two edges is likely connected to the 
morning-evening asymmetry seen in the high-phase mosaic. 
Returning to Fig.~\ref{hiphwac}, note that interior
to the ``blue" region of the E ring on the morning ansa, the E ring has a sharp inner edge that is not seen on the evening ansa. The relatively empty region interior to this edge probably corresponds to the region of reduced brightness on the shadow's morning edge  seen in the SHADBOUND data. This suggests that while the morning side of the E-ring has a higher particle density than the evening side  between 210,000 and 240,000 km (cf. Fig.~\ref{hpw175}), the opposite is true between  170,000 and 190,000 km. Unfortunately, due to difficulties in determining the instrumental background levels in the mosaics and complications associated with changing phase angles,  we cannot generate profiles like those in Fig.~\ref{hpw175} that clearly document this transition. Nevertheless, such a transition is a reasonable interpretation of  the combined data from the high-phase mosaics and the shadow-boundary observations. The location of this transition is interesting, because
it occurs near the location of the tip of Saturn's shadow in the planet's equatorial plane.
Together with the evidence for the sharp change in the E-ring's brightness between
the morning and evening sides of Saturn's shadow, this strongly suggests that 
the shadow influences the structure of the inner E ring.

Finally, note that on the shadow's evening side the sharp brightness transition deviates from the expected position of the shadow edge in Saturn's equator plane at radii greater than 190,000 km. As shown in Fig.~\ref{shadboundim} the observed position of the brightness boundary in this region would correspond to the expected
shadow edge at a location between 0 and 2000 km south of Saturn's equator plane. 
Thus the apparent position of the shadow boundary could reflect the inner E-ring's southward deflection observed in edge-on images (see Figs.~\ref{parplot2} and \ref{e105phasepar}). The proximity of the brightness edge to the predicted shadow edge at $z=0$ for radii less than 190,000 km would then imply that the ring is not vertically offset interior to 190,000 km. However, we should note that the contributions from Saturn's very flat G ring may be affecting the apparent position of
the shadow edge. Furthermore, we cannot completely rule out the possibility that the brightness variations seen outside 190,000 km do not trace the position of the shadow edge but instead reflect variations in the rings' surface density in the vicinity of the shadow. 

\section{Synthesis}

The above observational data document several interesting structures and asymmetries in Saturn's E ring, which should help elucidate the dynamics of 
E-ring particles.  Unfortunately, the observed data only provide measurements
of the brightness density as a function of radius, longitude and $z$, and the 
mapping between these real-space coordinates  and particle orbital elements 
is not straightforward. In particular, there is often insufficient information to determine 
whether  a given radial brightness distribution is due to variations in either the particles'  semi-major axes or their eccentricities.  Over the years, numerical
simulations have been used to predict the spatial distribution of particles moving under the influence of a specified set of forces and compare those with the observed data \citep{Horanyi92, Juhasz02, Juhasz04, Juhasz07, Horanyi08}. However, a thorough comparison of these models with the data is beyond the scope of this paper because the relatively complex dynamics of E-ring particles make it difficult  to exhaustively explore the relevant parameter space. Instead, we will focus on a few features of the data where analytical expressions can clarify the physical processes involved.

First, we will consider the average vertical warp and the double-banded structure in the vicinity of Enceladus' orbit. Here, analytical calculations  demonstrate that these features of the E ring are consistent with current models and theories of the E ring and we can therefore use these features to constrain both the physical and orbital properties of the E-ring particles. Next we will discuss the azimuthal asymmetries, which are difficult to explain in terms of existing theory and models. Here we will highlight aspects of the ring's structure that will likely pose the biggest challenge for theoretical models or numerical simulations. 

 \subsection{Average vertical warp}
 \label{warp}
 
\begin{figure}[tbp]
\resizebox{6in}{!}{\includegraphics{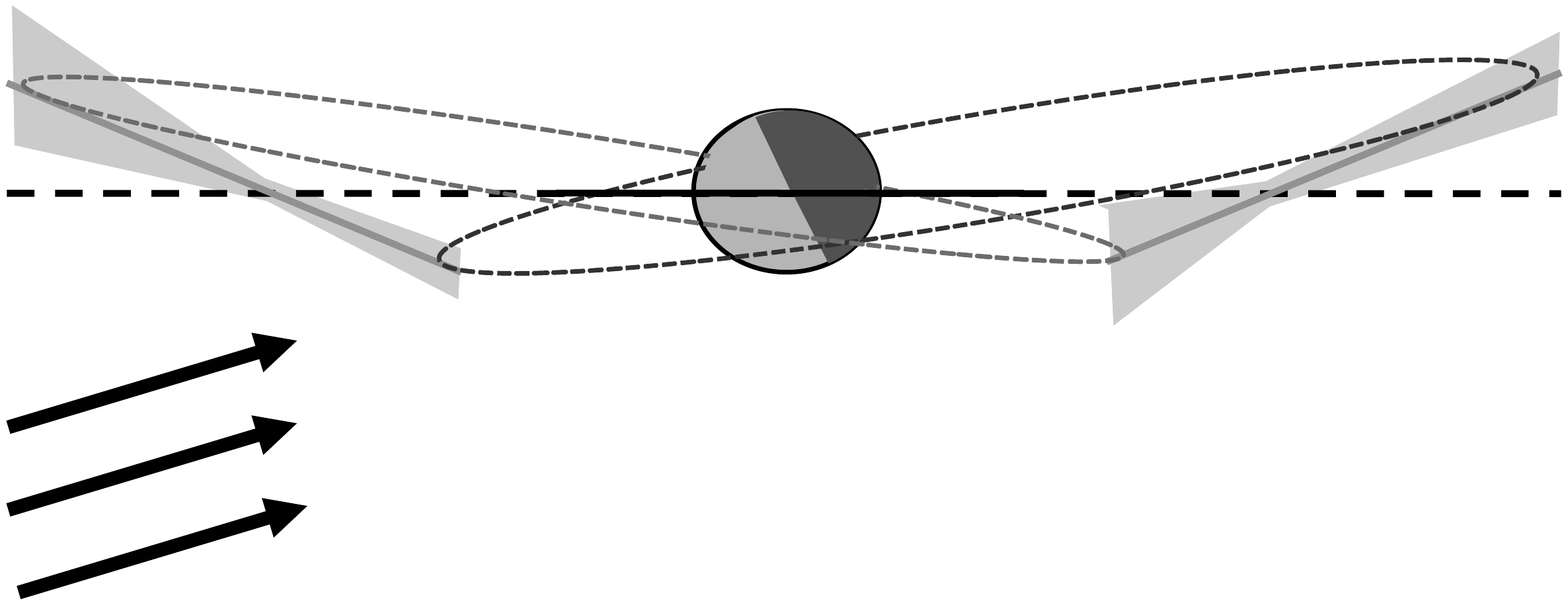}}
\caption{Sketch of the E-ring's vertical structure. The grey bands on either side
of Saturn illustrate the symmetric warp and asymmetric flare of the ring, while the
ellipses indicate representative orbits with arguments of pericenter near -90$^\circ$ (see Section 7.1). Note the planet and rings are not shown to scale.}
\label{eringdraw}
\end{figure}
 
\begin{figure}[tbp]
\resizebox{6in}{!}{\includegraphics{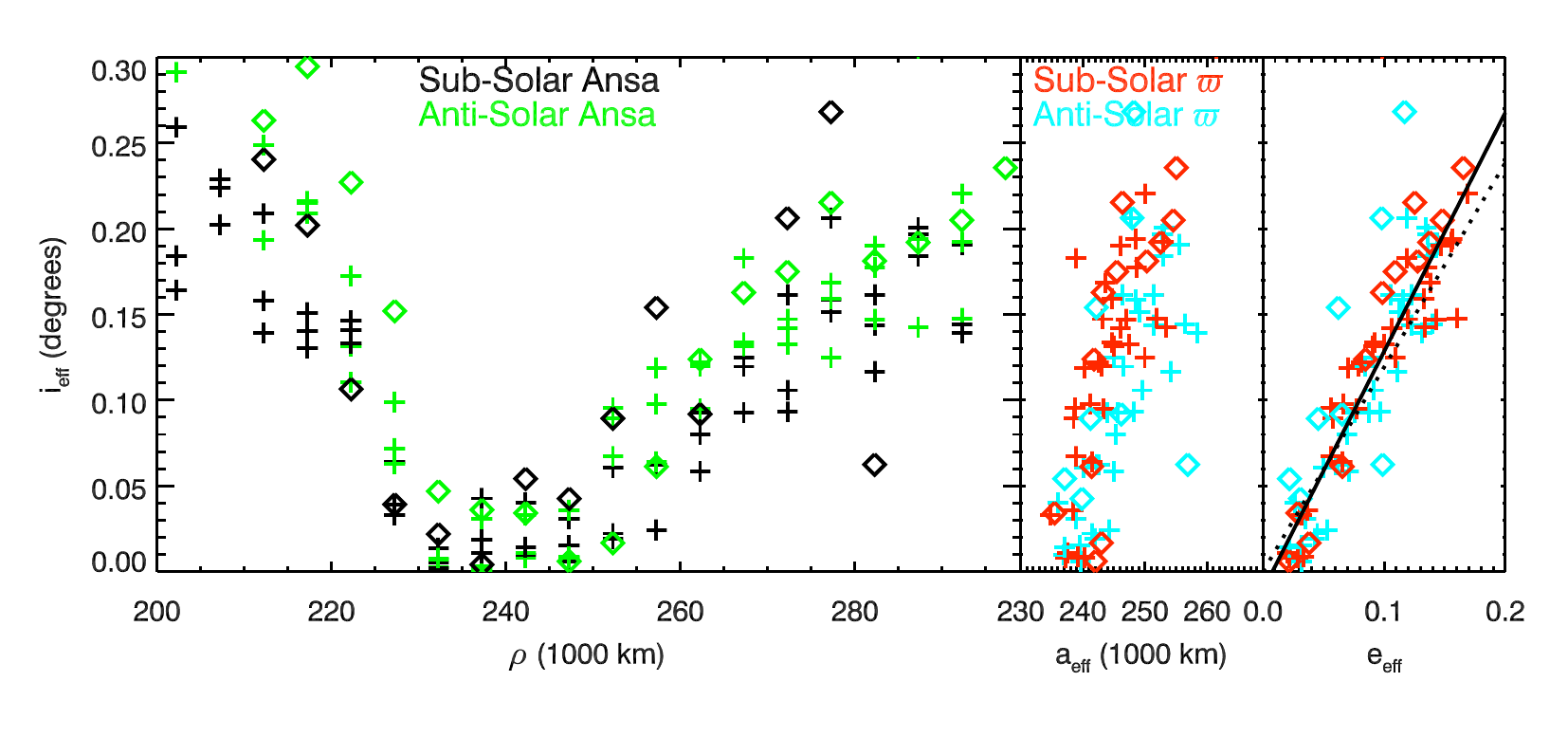}}
\caption{Effective orbital parameters derived from edge-on images of the E ring. 
Plus signs come from the E105PHASE observations (see Fig.~\ref{e105phasepar}), while the diamonds come from the E130MAP sequence (see Fig.~\ref{parplot2}). The left-hand panel 
shows the effective mean inclination of the ring particles as a 
function of radius in the ring.  Assuming that regions with similar effective inclination 
inward and outward of Enceladus correspond to particles on similar orbits
at periapse and apoapse, we compute the effective semi-major axis and effective
eccentricity  of particles as a function of inclination, shown in the right two panels
(In each case the periapse data from one ansa are matched to the
apoapse data on the opposite ansa.) The dotted and solid lines in the rightmost plots
show two different linear fits to all the data ($\sin i_{eff}=0.0208 e_{eff}$ and $\sin i_{eff}=0.0243 e_{eff}-0.0002$, respectively).}
\label{aeiplot}
\end{figure}
 
Edge-on observations at a variety of longitudes consistently show that the 
E-ring's peak brightness density  shifts from $\sim$1000 km southwards of 
Saturn's equator plane to $\sim$1000 km northwards of Saturn's equator plane 
between the orbits of Mimas and Tethys (see Figs.~\ref{parplot2}, ~\ref{e105phasepar}, and ~\ref{eringdraw}). The simplest interpretation of this shape is that  a significant fraction of the  E-ring particles in this region are on eccentric, inclined orbits with arguments of  pericenter (i.e. the angle between the orbit's ascending node and the orbit pericenter) $\omega$ around $-90^\circ$. For such orbits, the longitudes of their ascending nodes on Saturn's equatorial plane $\Omega$ are roughly 90$^\circ$ ahead of its longitude of pericenter $\varpi$, so these particles are always  found south of Saturn's equator plane near their orbital pericenters  and north of Saturn's equatorial plane near their orbital apocenters (see Fig.~\ref{eringdraw}). Note that  such a constraint
 on $\omega$ does not necessarily require $\Omega$ or $\varpi$ to have any particular value. Indeed, since the warp is observed at various longitudes, the E-ring must contain particles with all possible orbital pericenter longitudes or, equivalently, ascending nodes (see also below). 
 
 This distribution of orbital elements is not entirely unexpected. Numerical studies of eccentric circumplanetary orbits previously demonstrated that non-gravitational forces can cause $\omega$ to become locked around  $\pm90^\circ$ \citep{Horanyi90, Horanyi92, Hamilton93, Juhasz04}. \citet{Hamilton93} provides analytical expressions that demonstrate how this locking can arise from the out-of-plane components of solar radiation pressure and the Lorentz forces from Saturn's magnetic field and derives the appropriate governing equations.  However, these new data permit us to examine these processes in more detail than was previously possible and to test this theoretical model.

As mentioned above, there is no completely
generic way to  transform an onion-peeled brightness distribution in $\rho$ and $z$ into
a distribution in orbital elements. However,  the presence of detectable vertical offsets enables us to
estimate the ``typical'' orbital elements of the E-ring particles
at various locations within the rings. Let us assume that most of the E-ring particles between the orbits of Mimas and Tethys are on eccentric, inclined orbits with $\omega\simeq\pm90^\circ$, so the particles reach the extremes of their vertical motions at pericenter and apocenter. Since $d\rho/dt \simeq dz/dt \simeq 0$ at these locations, the time any particle spends in a small range of $\rho$ and $z$ will be a local maximum at its orbital pericenter and apocenter. Particles with a given semi-major axis $a$, eccentricity $e$ and inclination $i$ will therefore make the strongest  contributions to the ring's edge-on brightness density at the locations where $\rho=a(1\pm e)$ and $z=\pm\rho \sin i$ (cf. Fig. 3 of Horanyi {\it et al.} 1992). This suggests that the onion-peeled brightness of the ring at a given $\rho$ can often be dominated by particles with pericenters or apocenters at the specified value of $\rho$, in which case the vertical profile at that location would reflect the distribution of orbital inclinations of those particles. Of course particles spend a finite amount of time between periapse and apoapse, so the observed vertical distribution will be
contaminated at some level by particles with periapses less than $\rho$ and apoapses greater than $\rho$. Nevertheless, even  if we neglect these complications and assume that particles are only seen near their periapses and apoapses, we can still obtain illuminating information regarding the trends and correlations among the orbital elements of the E-ring particles.

Let us define a mean ``effective'' inclination of the particles at a given radial location $\rho$ as simply $\tan i_{eff}=|z_{off}|/\rho$, where $z_{off}$ is the measured vertical offset at that location shown in Figs.~\ref{parplot2} and~\ref{e105phasepar}. As shown in the left-hand panel of Fig.~\ref{aeiplot}, this effective mean inclination increases on either side of Enceladus' orbit, reaching roughly 0.2$^\circ$ around 200,000 km and 300,000 km. Given that Enceladus is the source of the E ring, we may assume that most of the particles  observed outside 240,000 km are near apoapse and those interior to 240,000 km are near periapse. Thus the data interior to 240,000 km show how the inclination varies with periapse distance, while the data exterior to 240,000 km show how the inclination varies with apoapse distance. Turning this around, we may regard the two radial locations where the ring has the same effective mean inclination as indicating the
``effective mean pericenter distance'' $a_{eff}(1-e_{eff})$ and ``effective mean apocenter distance'' $a_{eff}(1+e_{eff})$ of particles with those inclinations. These two numbers then yield an effective semi-major axis $a_{eff}$ and effective eccentricity $e_{eff}$ of the particles with each effective inclination. These parameters are plotted in the right-hand panels of Fig.~\ref{aeiplot}. Note that for these calculations we take the pericenter distances from observations of one ansa and the apocenter distances from a separate observation of  the opposite ansa. While there may be some slight differences in the trends derived from opposite ring ansa, they are fairly subtle and will not be considered further here.

Despite their crudity, the above calculations reveal interesting correlations among the various orbital elements. The effective inclination and effective eccentricity are strongly correlated, which is perfectly sensible given that vertical forces are more efficient at tilting eccentric
orbits due to the differences in the torque applied at periapse and apoapse (see below). 
In addition, the effective semi-major axes are all between 0 and $\sim$10,000 km exterior Enceladus' orbit. This is reasonable as Enceladus is the E-ring's primary source, and small grains launched from that moon are expected to have semi-major
axes that are nearby but somewhat exterior to the moon's semi-major axis
due to the charging of the grains upon exposure to the magnetosphere \citep{SB87}. Furthermore,  the forces that perturb semi-major axes (like plasma drag and electromagnetic forces associated with Saturn's shadow) tend to cause the semi-major axis to slowly migrate outwards \citep{HB91, BHS01, Horanyi08, HK08}.

To further evaluate these trends and correlations, we may compare
the observed relationship between $e_{eff}$ and $i_{eff}$ with that expected due to the action
of various vertical forces. \citet{Hamilton93} derived Gauss perturbation equations for E-ring particles moving under the influence of Saturn's oblateness, solar radiation pressure and Lorentz forces from Saturn's offset dipolar magnetic field, but ignoring asymmetric terms  associated with Saturn's shadow, etc. If the particle's orbit has a sufficiently
slowly varying eccentricity $e$,  the equations of motion have a solution where the argument of pericenter $\omega$ and inclination $i$ are
(Eq. 33 of Hamilton 1993a):
\begin{equation}
\sin \omega_{eq}= {\rm sign}(\dot{\omega}_{xy}/Z),
\label{signw}
\end{equation}
\begin{equation}
\sin i_{eq}=|Z/{\dot{\omega}_{xy}}|,
\end{equation}
where ${\rm sign}(x)=x/|x|$, $\dot{\omega}_{xy}$ is the precession rate for the argument of periapse and
$Z$ is a rate determined by the out-of-plane forces acting on the particle. Assuming small inclinations and making a small formatting change from Eqs. 31-32 of Hamilton (1993a), these rates are:
\begin{equation}
\dot{\omega}_{xy}=n\left[\frac{3J_2R_p^2}{a^2(1-e^2)^2}
-\frac{L}{(1-e^2)^{3/2}}\left(1-e^2-\frac{3n}{\Omega_p}\right)
+\frac{\alpha(1-e^2)^{1/2}}{ne}\cos\phi_\Sun\right],
\label{omegaxy1}
\end{equation}
\begin{equation}
Z=\frac{ne}{(1-e^2)^{1/2}}\left[\frac{\alpha}{n}\sin B_\Sun +
\frac{3L}{2(1-e^2)^2}\left(\frac{g_{2,0}}{g_{1,0}}\right)\left(\frac{R_p}{a}\right)
\left(\frac{n}{\Omega_p}\right)\right],
\end{equation}
where $R_p=60,330$ km is the planet's radius,
$\Omega_p\sim1.7*10^{-4}/$s is the planet's rotation rate, $J_2\sim0.017$
is Saturn's quadrupole gravitational harmonic \citep{Jacobson06},
$g_{1,0}\sim21.2 \mu$T and $g_{2,0} \simeq1.5 \mu$T are the dipole
and aligned quadrupole Gauss coeffcients of Saturn's magnetic field \citep{Gombosi09},
$a$ is the particle's orbital semi-major axis, $n$ is the particle's orbital 
mean motion, $\phi_\Sun$ is the angle between the particle's orbital 
pericenter and the subsolar longitude, $B_\sun$ is the solar elevation
angle, and $\alpha/n$ and $L$ are unitless force ratios. The ratio
$\alpha/n$ is set by the ratio of solar radiation pressure to the planet's 
gravity, and is:
\begin{equation}
\frac{\alpha}{n}=\beta\frac{3M_\Sun a^2}{2M_p a_p^2},
\label{alpha}
\end{equation}
where $M_\Sun=2*10^{30}$ kg is the Sun's mass, $M_p=5.7*10^{26}$ kg is Saturn's mass,
$a_p\sim 1.5*10^{9}$ km is the planet's heliocentric orbital semi-major axis, and $\beta$ is the ratio of the solar radiation force to the solar gravitational
force acting on the particle, which depends on the particle's physical and 
optical properties \citep{BLS79}. The parameter $L$ is a measure
of the strength of the electromagnetic force relative to the planet's gravity and
is given by (modified to mks units and corrected to fix a typographical error 
from Equation 22 in Hamilton 1993a, see Hamilton 1993b): \nocite{Hamilton93b}
\begin{equation}
L=\frac{q_g}{m_g}\frac{g_{1,0}R_p^3\Omega_p}{GM_p},
\end{equation}
where $q_g$ and $m_g$ are the grain's charge and mass. The
charge-to-mass ratio can be expressed as $3\epsilon_o \Phi/ \rho_g s^2$, where
$\epsilon_0$ is the permittivity of free space, $\Phi$ is grain's electrostatic
potential, $\rho_g$ is the grain's mass density, and $s$ is the grain radius.

In order to evaluate $Z$ and $\dot{\omega}_{xy}$ numerically, we will assume that the particles have semi-major axes close to Enceladus' ($a\simeq 240,000$ km  and $n \simeq 5.2*10^{-5}/$s) and have moderate eccentricities ($e<0.25$), consistent with the trends seen in Fig.~\ref{aeiplot}.
In this case, we may approximate terms like $1-e^2$ as simply unity. We also assume  $B_\sun \simeq -17^\circ$, which is the average solar ring opening angle during these observations. Also, as discussed in Section 2, the particles that dominate the observed E-ring brightness in the Cassini images should be of order 1 $\mu$m in radius. Furthermore,  various observations indicate that E-ring particles are composed  primarily of water-ice  and that the largest particles are charged to roughly  -2 V \citep {Kempf06, Hillier07, Postberg08, Hedman09}. Hence the parameter $\beta\sim1$ \citep{BLS79} and we may express the charge-to-mass ratio  $q_g/m_g \sim -0.05$ C/kg $(\Phi/$-2V$)/(s/1\mu$m$)^2$ (assuming $\rho_g = 1$ g/cm$^3$, which may be an overestimate if the grains have an aggregate or porous structure).  Thus the unitless parameters are $\alpha/n\simeq1.3*10^{-4}\beta$ and 
$L\simeq-9.6*10^{-4} (\Phi/$-2V$)/(s/1\mu$m$)^2$ and:
\begin{equation}
\dot{\omega}_{xy}=
n\left[0.0032+0.00005\frac{\Phi/(-2V)}{(s/1\mu m)^2}\left(1-\frac{e^2}{0.05}\right)+0.00013\beta \frac{\cos\phi_\Sun}{e}\right],
\label{prxy}
\end{equation}
\begin{equation}
Z=-ne\left[3.9*10^{-5} \beta + 0.8*10^{-5}\frac{\Phi/(-2V)}{(s/1\mu m)^2}\right].
\end{equation}
The last term in Eq.~\ref{prxy} is negligible unless $e$ is much less 
than 0.1, and the second term is positive for  $e<0.23$. Thus
for the particles of interest here, $\dot{\omega}_{xy}$  and $Z$ have opposite signs, which means (from Eq.~\ref{signw}) $\sin \omega =-1$ or $\omega=-90^\circ$,
as observed. Also, the equilibrium inclination is given by (again, assuming $1-e^2 \simeq 1$):
\begin{equation}
\sin i_{eq} \simeq 0.012 e \left[\beta + 0.2\frac{\Phi/(-2V)}{(s/1\mu m)^2}\right].
\end{equation}
Fitting a line through the estimated  effective inclinations and eccentricities
in Fig.~\ref{aeiplot}, we obtain $\sin i_{eff} \sim 0.021 e_{eff}$. Such a trend
is consistent with the above expression if $\beta \sim 1$ and $s \sim 0.5 \mu$m.
Of course, $s$ could be somewhat larger if the grains have significant porosity or fractal structure (such that $\rho_g<1$ g/cm$^3$) or smaller if the observed grains are less negatively charged than the larger particles observed by the in-situ experiments. Nevertheless, it is remarkable that the required $s$ is close  to the particle size that is expected to dominate the appearance of the E ring in the Cassini images (see Section 2 above). Given the roughness of  these calculations, this level of agreement encourages us to regard the effective orbital elements derived above as reflecting real properties of the E ring. Furthermore, this analysis also confirms that the vertical component of solar radiation pressure can produce sufficiently high inclinations to generate the observed  large-scale warp.

\subsection{Vertical structure near Enceladus' orbit}

Between 230,000 and 280,000 km from Saturn's center the E-ring's vertical density profile departs significantly from the simple Lorentzian form found elsewhere in the ring (see Figure~\ref{parplot2}).  Instead, the observed vertical profiles in this region are best fit by a broad (4000-5000 km FWHM)  Lorentzian peak plus a narrow ($\sim$2000 km FWHM) Gaussian dip. This distinctive vertical structure is also present in many numerical simulations \citep{Juhasz07, Kempf10, Agarwal11}, and has also been seen in RPWS data \citep{Kurth06}. However, this structure is  not clearly visible in the HRD data from Cassini's Cosmic Dust Analyzer \citep{Kempf08, Kempf10}, 
which may be in part due to differences in the range of particle sizes  probed by these observations (cf. Juhasz {\it et al.} 2007). This pattern is almost certainly due to a combination of the non-zero speeds at which particles are launched from Enceladus' surface and the particles' gravitational scattering during close encounters with the moon. These phenomena  excite the particles' vertical velocities, reducing the local particle density near the ring-plane over a range in $z$ comparable in size to Enceladus'  Hill sphere. The observed radial trends  in the vertical structure are indeed compatible with a structure produced by close Encealdus encounters. 

The magnitude of the mid-plane density depletion (measured using the ratio of the Gaussian and Lorentzian terms in the fitted profiles --see  the second panel of Fig.~\ref{parplot2}) clearly peaks around Enceladus' semi-major axis $a_E$, which implies that Enceladus plays an important role in producing this structure. However, we also need to explain the asymmetric profile of the mid-plane density depletion, which extends much further exterior to $a_E$. Recalling that a particle on an eccentric orbit contributes most strongly to the ring's brightness near its orbital pericenters and apocenters (cf. Figure 3 in Horanyi {\em et al.} 1992), the simplest interpretation of this asymmetry is that the affected ring particles all have pericenters close to $a_E$, and a range of apocenter distances extending between $a_E$ and $ a_E+30,000$ km, The affected particles therefore have semi-major axes between $ a_E$ and $a_E+15,000$ km, comparable to the range of semi-major axes derived from the E-ring's vertical warp (see Fig.~\ref{aeiplot}). As mentioned above, this range of semi-major axes is likely due to the charging of the small grains upon escaping from Enceladus \citep{SB87}, effects from the planetary shadow edge \citep{HB91, HK08}, as well as slow outward migration due to various non-gravitational forces \citep{BHS01, Horanyi08}. However, the eccentricities of these particles are all less than $\sim 0.06$, which is much less than the range of effective eccentricities found in Fig.~\ref{aeiplot}. 

Both the pericenter distances and eccentricities of the particles involved in the mid-plane density depletion can be explained in terms of the dynamics of close Enceladus encounters.  For particles with semi-major axes greater than $a_E$, the likelihood of a close encounter is greatly enhanced for particles whose pericenters are close to $a_E$ because $d\rho/dt \simeq 0$ at those locations. Furthermore, in order for Enceladus  to significantly perturb a particle's orbit, the velocity of the particle relative to the moon during the encounter $v_{rel}$ cannot be much larger than the Enceladus' nominal escape speed $v_{esc}\sim0.24$ km/s \citep{Kempf10, Agarwal11}. A particle with eccentricity $e$ encountering Enceladus at periapse will pass the moon at a $v_{rel} \simeq (\sqrt{1+e}-1)*v_E$, where $v_E\simeq 12$ km/s is the orbital speed of Enceladus. Thus $v_{rel} \simeq v_{esc}$ when the particles' orbital eccentricity is about 0.04, comparable to the maximum observed eccentricity of the particles involved in the mid-plane density depletion. We can therefore conclude that inclination excitation by close encounters with Enceladus is a reasonable explanation for the vertical structure of the E ring between 230,000 and 280,000 km.

\subsection {Hour-angle variations}

Both the E-ring's vertical thickness and its vertically-integrated brightness profile vary with longitude relative to the Sun. Again, this result was not entirely unexpected because azimuthal asymmetries have been observed in numerical simulations of diffuse rings perturbed by asymmetric forcing terms like solar radiation pressure, changes in the ambient plasma or charge fluctuations generated by the particles' passage through the planet's shadow  \citep{BS89, HB91, Hamilton93, BHS01, Juhasz04, HK08}. However, several of the observed brightness asymmetries are very different from those found in published models, and consequently are difficult to understand and interpret.

The brightness asymmetries between the sub-solar and anti-solar sides of the E ring are particularly problematic. Figures~\ref{e105phasepar} and~\ref{hpw177} both indicate that the entire inner part of the E-ring (between 180,000 and 250,000 km) is roughly 50\% brighter on the ring's sub-solar side than it is on the anti-solar side, while outside this region the two sides of the ring have nearly the same brightness. In both cases, the two sides of the ring were observed at nearly the same viewing geometry, so this difference cannot be a photometric effect. Instead, it appears that the total amount of material visible on the sub-solar ansa is higher than the total amount of material visible on the anti-solar ansa.  This is a surprising result because while asymmetric forces like solar radiation pressure can generate differences in the radial distribution of material on opposite sides of the planet \citep{BHS01, Juhasz04, HK08}, the ring should still consist of particles in orbit around the planet, so the total amount of material visible on both sides should be nearly the same (after accounting for variations in the particles' orbital speed with true anomaly). 

This asymmetry becomes even more puzzling if we also take the ring's  vertical warp into account. The above analysis of this warp indicates that most of the E-ring particles observed between 180,000 and 300,000 km have semi-major  axes close to $a_E=240,000$ km. If this is correct, then the excess brightness between 180,000 and 240,000 km on the sunward ansa would be associated with particles near their orbital periapses. However, these same particles would be near their apoapses on the anti-solar side of the planet, so we would expect the anti-solar ansa to be brighter than the sub-solar ansa between 240,000 and 300,000 km, which is not the case (the brightness profiles are actually nearly identical exterior to 250,000 km). The relevant particles also probably cannot be hidden within Saturn's shadow or in a region far beyond 300,000 km, because this would produce systematic differences in the mean vertical offset profiles between the two ansa that are incompatible with the observations. If the particles responsible for the brightness excess had apoapses much less than 250,000 km or much greater than 300,000 km, then they would have significantly different semi-major axes and eccentricities from the particles with periapses on the ring's shadowed side. This would in turn alter the inclinations induced by the out-of-plane forces (see Section~\ref{warp} above), yielding a different southward vertical offset on the sub-solar ansa than the anti-solar ansa, which is not observed (see Fig.~\ref{e105phasepar}).  Thus reconciling these observations will pose a significant challenge to any theoretical and numerical models of the E ring.

In the absence of an obvious simple model for the E-ring's enhanced brightness near the sub-solar longitude, we will simply suggest some possible explanations that are worthy of future exploration. The most straightforward option is that there is some non-trivial correlation between particles' eccentricities and pericenters that  crowd streamline trajectories between 180,000 and 250,000 km on the sunward side of  the rings, and disperse the relevant material over a wide radius range at other longitudes, making the brightness excess at anti-solar longitudes difficult to detect. We could also consider more exotic scenarios, such as  hour-angle variations in the individual ring particles' light-scattering properties, or the particles not following the expected Keplerian orbits due to some interaction with the magnetospheric plasma. 

The remaining hour-angle asymmetries in the ring are less paradoxical. For example, unlike the sub-solar/anti-solar asymmetries discussed above, the differences between the morning and evening sides of the ring seen in VIO-filter data (see Fig.~\ref{hpw175}) probably can be attributed to  differences in the eccentricity distributions of particles with pericenters on the morning and evening sides of the planet. Assuming that the brightness at a given radius is dominated by particles near their  orbital pericenters or apocenters, then the shape of the brightness  profile interior to Encealdus  should reflect the eccentricity  distribution of particles with longitudes of pericenter  $\varpi$ near the observed longitude. The morning ansa's profile is brighter that the evening ansa between 200,000 and 240,000 km (Fig.~\ref{hpw175}), and fainter interior to 200,000 km (Fig.~\ref{shadboundprof}). Thus the particles with $\varpi \simeq \lambda_\Sun-90^\circ$ would appear to have an excess of low-eccentricity orbits and a depletion of high-eccentricity orbits compared  to the particles with $\varpi \simeq \lambda_\Sun+90^\circ$. Support for this interpretation of  the data  can be obtained by examining  the brightness profiles exterior to Enceladus' orbit, where the relevant particles should reach their apoapses. In particular,  the material that causes the morning ansa to appear brighter between 210,000 and 240,000 km should also cause the evening side of the rings to appear brighter between 250,000 km and 280,000 km, and such a brightness excess  may be present in the dark blue VIO-filter profile of Fig.~\ref{hpw175}. 

The position of the sharp edge in the morning ansa's brightness lies around 200,000 km, near the location of the tip of Saturn's shadow in the $z=0$ plane (see Figs.~\ref{hiphwac} and ~\ref{shadboundim}). Thus the ring's brightness profile on the morning side of the planet and the lack of material on the morning edge of the planet's shadow could  indicate that the particles with sufficiently high eccentricities to enter the shadow near periapse had their eccentricities reduced and their periapses raised until their orbits no longer intercepted the shadow. If we also consider the relevant orbital precession rates, such a scenario can even explain why the particles with periapses on the two sides of the ring have different eccentricity distributions.

If we neglect the effects of solar radiation pressure, then using Eqs. 30-31 of  \citet{Hamilton93} we can estimate the apsidal precession rate of the E-ring particles as:
\begin{equation}
\dot{\varpi}_x=n\left[\frac{3J_2R_P^2}{2a^2(1-e^2)^2}
	+\frac{2nL}{\Omega_p(1-e^2)^{3/2}}\right]
\end{equation}
(Note this expression differs from that given in Eq.~\ref{omegaxy1} because it gives the precession rate of the pericenter longitude relative to an inertial direction instead of relative to the ascending node.) This equation is evaluated using the same parameters as discussed in Section~\ref{warp} above to yield:
\begin{equation}
\dot{\varpi}_x=0.42^\circ/day\left[1-0.39\frac{\Phi/(-2V)}{(s/1\mu m)^2}\right].
\label{precx}
\end{equation}	
For these nominal values $\dot{\varpi}_x>0$ and it takes about 4 years for the particle's pericenter to drift  all the way around the planet (in contrast to the simulations of Hamilton 1993a, which assumed a larger nominal charge-to-mass ratio, so  $\dot{\varpi}_x<0$). Thus particles with pericenters on the morning ansa had their periapses aligned with Saturn's shadow less than a year ago,  while the particles with periapses on the evening ansa last had their pericenters aligned with Saturn's shadow over 3 years ago. The difference in timing between these ``periapse shadow passages'' is significant, because during the few years before these observations, the tip of Saturn's shadow has been gradually moving outward into the The tip of Saturn's shadow only reached into the E ring between 2005 and 2006, so E-ring particles with $\varpi \sim \lambda_\sun-90^\circ$ could have encountered Saturn's shadow during their last periapse shadow passage, while E-ring particles with $\varpi \sim \lambda_\sun+90^\circ$ would not have been able to encounter the shadow during their last periapse shadow passage. Hence, only particles with periapses on the morning side of the rings would be likely to have their orbits altered by passage through Saturn's shadow, supporting the above notion that such shadow passages
could have raised the relevant particle's orbital pericenters.

While these calculations are suggestive, it is important to note that the precession rate is sensitive to particle size, and the pericenter will regress instead of precess if $s<0.7 \mu$m.  Such particle-size-dependent variations in the precession rate can have important effects on the spatial distribution of E-ring particles \citep{HB94, Horanyi08}. Even though exploring these phenomena in detail is beyond the scope of this paper, we can at least point out that the particles cannot be too small if the precession rate is to be sufficiently large and positive for shadow-induced perturbation to be a plausible explanation for the morning-evening asymmetries. At the same time, the particles cannot be too large or else the expected non-gravitational forces would not be  able to produce the observed large-scale warp in the ring (see above). Fortunately, the cameras are most sensitive to particles in the 0.5-1.0 $\mu$m size range, which is roughly compatible with both these constraints. This bodes well for future efforts to model these features, but it does suggest that there may only be a rather restricted range of particle parameters that will be consistent with all the observed E-ring structures. 

Another asymmetry that may be associated with Saturn's shadow is the difference in the E-ring's vertical thickness between the sub-solar and anti-solar sides of the planet.  As shown in Figs.~\ref{e105phasepar} and~\ref{eringdraw}, the ring is thicker on the anti-solar side of the rings at radii interior to $a_E$, and is thicker on the sub-solar side at larger radii. This suggests that particles with periapses on the planet's shadowed side have a broader range of inclinations than those with periapses on the planet's sunward side. So long as $|\dot{\varpi}_x|$ is less than about 0.5$^0/$day, particles with periapses on the sunward side of the planet in 2006 would not have encountered the shadow during their previous periapse shadow passage, while the particles with periapses on the anti-solar side of the planet would have entered the shadow. Thus only the latter particles are likely to experience shadow-induced perturbations, which may have excited inclinations and produced a thicker ring. 

Despite the above evidence for the shadow's role in generating these asymmetries, the precise physical processes involved remain unclear and require further examination. For example, the ring's brightness profile on the morning side of the planet and the lack of material on the morning edge of the planet's shadow both suggest that particles with sufficiently high eccentricities to enter the shadow near periapse had their eccentricities reduced and their periapses raised until their orbits no longer intercepted the shadow. However, published theoretical studies of the orbital perturbations induced by a planet's shadow have focused on the reduction in solar radiation pressure and the Lorentz forces associated with charge variations \citep{HB91, Juhasz04, HK08}, both of which are primarily {\em radial} perturbations
that cannot efficiently reduce orbital eccentricities when applied near periapse \citep{Burns76}. Thus more complex interactions between the ring particles and the plasma environment in the vicinity of Saturn's shadow may need to be considered. Indeed, recent plasma measurements within the core of the E ring may imply that  dusty plasma phenomena could be relevant to the charge state of the E-ring particles \citep{Wahlund09}.

The above discussions demonstrate that more work needs to be done to identify the forces responsible for the observed hour-angle asymmetries in the E ring. However, a fully self-consistent model of the ring will not only have to reproduce the observed asymmetries, but also explain why the expected asymmetries from well-studied non-gravitational perturbations such as solar radiation pressure are not obvious in the observed data. For example, including solar radiation pressure introduces an asymmetry into the equations of motion that prevents particles' orbits from maintaining a fixed eccentricity as they precess around the planet. Instead, there is a specific solution  with a finite ``forced eccentricity'' $e_f$ that maintains a fixed orientation with respect to the Sun, and in general the particles' orbits oscillate around this steady-state solution \citep{Hedman10}. Using Eqs. 27, 31 and 32 in \citet{Hamilton93} and again approximating $1-e^2$ as unity (see above), we can derive the eccentricity $e_f$ and longitude of pericenter $\varpi_f$ of this steady-state solution:
\begin{equation}
\cos(\varpi_f-\lambda_\sun)=-\rm{sign}(\dot{\varpi}_x)
\end{equation}
and
\begin{equation}
e_f=\left|\frac{\alpha}{\dot{\varpi_x}}\right|.
\end{equation}
where $\alpha$ and $\varpi_x$ are given by Eqs.~\ref{alpha} and ~\ref{precx}, respectively. As before, inserting the numerical values for the relevant parameters used in  Section~\ref{warp} above,  we find that the forced eccentricity is:
\begin{equation}
e_f=0.081\beta\left|1-0.39\frac{\Phi/(-2V)}{(s/1\mu m)^2}\right|^{-1}.
\end{equation} 
Regardless of the sign of $\dot{\varpi}_x$,  such a large $e_f$ means that the various particles' orbits will have either pericenters that remain on one side of the planet or  substantially different eccentricities when the pericenters are on the sub-solar and anti-solar sides of the ring. Any model for the brightness asymmetries between
the sub-solar and anti-solar sides of the rings will therefore have to account for this forced eccentricity. 

\section{Summary and conclusions}

Images obtained in 2005-2006 have both confirmed the existence
of previously detected E-ring structures, and have also revealed
several new features. The E-ring structural properties found in this
analysis include:
\begin{itemize}
\item
The vertical particle density profile of the ring at radii greater than
280,000 km and less than 230,000 km can be best fit by a 
Lorentzian function.
 \item Between 230,000 and 280,000 km, there is a 
 localized depletion in brightness density relative to the
 best-fit Lorentzian profile within $\sim1000$ km
 of Saturn's equator plane. 
 \item 
The ring's peak brightness density shifts from roughly 1000 km
south of Saturn's equator plane to roughly 1000 km north of
Saturn's equator plane between the orbits of Mimas and Tethys.
\item 
The ring's vertical thickness increases with distance from Enceladus'
semi-major axis.
\item
The ring's overall brightness and vertical structure do not
vary significantly with longitude relative to Enceladus.
\item
The ring's vertical thickness varies with longitude relative to the Sun,
being thicker on the anti-solar side of the planet interior to Enceladus' semi-major axis and thicker on the sub-solar side further from Saturn.
\item 
The mean vertical offsets in the ring's peak brightness density
do not vary with longitude relative to the Sun.
\item
The ring's brightness profiles on the morning and evening sides
of the planet differ interior to Enceladus' semi-major axis, 
with the morning profile showing a sharp edge
around 200,000 km.
\item
The ring's brightness profiles on the sub-solar and anti-solar sides
of the planet differ interior to Enceladus' semi-major axis, 
with the sub-solar side being intrinsically brighter.
\end{itemize}

We have only just begun efforts to interpret these features in terms
of particle dynamics. The overall vertical warp can be reasonably 
attributed to the vertical components of solar radiation pressure
and Lorentz forces from Saturn's offset dipole field,  and the particle depletion 
close to the orbit of Enceladus is consistent
with gravitational perturbations from that moon. 
However, the various hour-angle asymmetries are still difficult  to
understand, so additional work will be 
needed in order to fully understand the azimuthal structure of this ring.

\section*{Acknowledgments}
We acknowledge the support of NASA, the Cassini project and the Cassini Imaging Team. This work was supported by grants from the Cassini Data Analysis, Planetary Geology and Geophysics, and Outer Planets Research programs. We also thank two anonymous reviewers for their helpful comments on an earlier draft of this manuscript.


\end{document}